\documentclass[a4paper]{article}

\usepackage[english]{babel}
\usepackage[utf8]{inputenc}
\usepackage[T1]{fontenc}
\usepackage[a4paper,top=3cm,bottom=2cm,left=3cm,right=3cm,marginparwidth=1.75cm]{geometry}

\usepackage{float}
\usepackage{subfig}
\usepackage{adjustbox}
\usepackage{amsmath}
\usepackage{graphicx}
\usepackage[colorinlistoftodos]{todonotes}
\usepackage[colorlinks=true, allcolors=blue]{hyperref}
\usepackage{authblk}
\usepackage[ruled,vlined]{algorithm2e}
\usepackage[left]{lineno}
\usepackage{csquotes}
\usepackage[numbers]{natbib}

\providecommand{\keywords}[1]{\textbf{\textit{Keywords:}} #1}

\title{An Adaptive Quasi-Continuum Approach for Modeling Fracture in Networked Materials: Application to Modeling of Polymer Networks}
\author[1]{Ahmed Ghareeb}
\author[1]{Ahmed Elbanna\footnote{Corresponding author: Ahmed Elbanna, elbanna2@illinois.edu}}
\affil[1]{Department of Civil and Environmental Engineering, University of Illinois at Urbana-Champaign, Illinois, USA}
\date{May 2019}

\begin{document}
\maketitle

\begin{abstract}
\noindent Materials with network-like microstructure, including polymers, are the backbone for many natural and human-made materials such as gels, biological tissues, metamaterials, and rubbers. Fracture processes in these networked materials are intrinsically multiscale, and it is computationally prohibitive to adopt a fully discrete approach for large scale systems. To overcome such a challenge, we introduce an adaptive numerical algorithm for modeling fracture in this class of materials, with a primary application to polymer networks, using an extended version of the Quasi-Continuum method that accounts for both material and geometric nonlinearities. In regions of high interest, for example near crack tips, explicit representation of the local topology is retained where each polymer chain is idealized using the worm like chain model. Away from these imperfections, the degrees of freedom are limited to a fraction of the network nodes and the network structure is computationally homogenized, using the micro-macro energy consistency condition, to yield an anisotropic material tensor consistent with the underlying network structure. A nonlinear finite element framework including both material and geometric nonlinearities is used to solve the system where dynamic adaptivity allows transition between the continuum and discrete scales. The method enables accurate modelling of crack propagation without a priori constraint on the fracture energy while maintaining the influence of large-scale elastic loading in the bulk. We demonstrate the accuracy and efficiency of the method by applying it to study the fracture in different examples of network structures. We further use the method to investigate the effects of network topology and disorder on its fracture characteristics. We discuss the implications of our method for multiscale analysis of fracture in networked material as they arise in different applications in biology and engineering.
\end{abstract}
\hspace{0.75 cm} \keywords{Quasi-Continuum Method, Polymer Networks, Fracture}

\section{Introduction}
\label{S:1}
Polymers are the building blocks for many natural and artificial materials. The load transfer system in polymeric materials may be abstracted as a complex network of cross-linked nonlinear  chains \cite{Creton2016FractureReview}. The topological properties of polymer networks, such as local connectivity, cross-linking density, bond types, and sacrificial bonds and hidden length, may dramatically affect their mechanical response and fracture proprieties \cite{Everaers2009TopologyNetworks,Kothari2018MechanicalAccumulation}. Most current experimental techniques do not allow direct visualization of the deformation and damage on the network micro scale, thus, the relation between the topology and mechanical behavior of these materials remains elusive. Multiscale numerical models are thus needed to bridge this gap \cite{Everaers2009TopologyNetworks}.

Modeling of soft polymeric materials has been mainly approached using continuum theories, including linear elasticity, hyperelasticity, viscoelasticity, viscoplasticity and poroelasticity \cite{doi:10.5254/1.3538357,Rivlin1948LargeTheory,M.Arruda1993AMaterials,Hu2010UsingGelsb,Chester2011AGels}. Continuum representation of materials has been used in traditional solid mechanics, where the material is assumed to be a continuous medium with its response described mathematically by a set of constitutive laws. Through experiments, the parameters of the constitutive laws may be determined and then utilized to solve the boundary value problems using either analytical or computational approaches. These techniques enabled significant progress in modeling the elasticity and distributed damage behavior of polymeric materials with and without infused fluids. However, these continuum methods are usually independent of the size scale of the network structures and they may not capture localized phenomena without enrichment, which in many cases is either expensive or is based on idealized models of microstructure. In particular, when it comes to fracture, a phenomenon that depends critically on the local conditions in the vicinity of the propagating cracks, continuum methods are not suited for predicting the effect of the network local topology on its fracture resistance. Furthermore, continuum methods may also fall short in guiding the network geometric design for enhanced fracture toughness.

An exception in the recent years has been the emergence of phase field models which is increasingly used to simulate fracture in a wide class of materials including gels and rubbers \cite{Mao2018FractureFailure,Miehe2014PhaseFailure,Kumar2018FractureImplementation}. The method may be used to simulate curved crack paths, crack kinking and branching, and crack front segmentation. Phase filed methods are based on gradient enhanced damage models where a parameter is introduced to distinguish between fully intact material and fully damaged material in a gradual manner \cite{Spatschek2011PhasePropagation}. The width of the damage zone depends on a length scale that is introduced primarily for regularization purposes but has been shown theoretically that as this length scale goes to zero, Griffith fracture criterion \cite{Griffith1921TheSolids} is recovered. The damage variable changes from 1 (fully intact) to 0 (Fully damaged) over this length scale. This length scale must be carefully chosen to regularize the strain-softening behavior during the fracture process, and to avoid mesh dependency-related issues during finite element simulations. The choice of this lengths scale for materials with microstructure, which possess physical length scales of their own, is not clear apriori and is left, in most cases, as a tunable parameter to fit experimental observations \cite{Zhang2017DeterminingExperiments}. 

Recently, thanks to advances in imaging, computational power, and nano/micro fabrication technologies, the mechanics of materials takes on another challenge of modeling fundamental material behavior starting from the atomic scale \cite{Buehler2008AtomisticFailure}. For the accurate representation of the material response at this scale, the material must be modeled using discrete simulations. \textcolor{black}{Discrete simulations have been widely used to model polymeric materials as networks of irregular lattices with different link models that incorporate the polymer chain physics \cite{Broedersz2014ModelingNetworks}. Adapting these discrete models where each polymer chain is modeled explicitly, it is possible to accurately describe localized phenomena such as fracture and cavitation. Furthermore, a new synthesis paradigm has been introduced in recent years for realizing near ideal polymer networks. These ideal covalent networks, which are produced by the cross-coupling of macro-molecular building blocks \cite{Sakai2008DesignMacromonomers,Sakai2010HighlyPolymers}, are essentially free of entanglements and have few topological defects which leads to high strength \cite{Sakai2008DesignMacromonomers}. Ideal networks provide a pathway for exploring an explicit connection between network design and its mechanical response \cite{Akagi2013UltimateStructure,Alame2019RelativeNetworks}, as well as a promising platform for designing of new lattice-like materials with tunable properties \cite{Kamata2014NonswellableHysteresis}. Motivated by these ideal networks, Alame and Brassart \cite{Alame2019RelativeNetworks} studied the elasticity of discrete near-ideal polymer networks, using a force extension model for polymers confined to a surface, with different coordination numbers. However, the study was limited to small scale samples for computational considerations.} 

In this paper, We introduce a new adaptive numerical algorithm for modeling fracture in polymer networks using an extended version of the Quasi-Continuum method \cite{Tadmor1996QuasicontinuumSolids,Miller2002,Shenoy1999AnMethod} that accounts for both the nonlinear elastic nature of the polymer chains as well as the geometric nonlinearity associated with their potentially large deformation. In regions of high interest where deformation gradients are non uniform, for example near crack tips, explicit representation of the local topology is retained where each polymer chain is idealized using the worm like chain model. Away from these imperfections where the deformation gradient is sufficiently uniform, the network structure is computationally homogenized, using an equivalence of the microscale and macroscale incremental work, to yield an anisotropic material tensor consistent with the underlying network structure. Thus at any instant in time, only a fraction of the network nodes is solved. Dynamic adaptivity allows efficient transition between the two resolutions. Our goal is to develop a method that enables modeling crack propagation without apriori constraint on the fracture energy or the need to assume phenomenological length scales, while maintaining the influence of large-scale elastic loading in the bulk. 

The remainder of the paper is organized as follows. In Section \ref{sec:overview} we present an overview of the QC method. We introduce the problem formulation, the homogenization technique, and the numerical implementation algorithms used in this study in Sections 3 and 4. We verify the numerical algorithm for modeling both pristine and cracked samples in Section 5. In Section \ref{sec:network}, we demonstrate the efficiency of the method by applying it to study the fracture of in model systems that are prohibitive for a full-field discrete approach. We further use the method to study the effects of network topology on its fracture resistance. Finally, We discuss the method implications for the analysis of networked material systems in Section \ref{sec:discussion}.

\section{Quasi-Continuum method overview:}\label{sec:overview}
\label{S:2}
Tadmor, Ortiz, and Phillips \cite{Tadmor1996QuasicontinuumSolids} proposed the Quasi-continuum (QC) method in 1996 to overcome the scale limitations of atomic simulations. The original QC method is a computational technique to model an atomistic system without the need to treat all atoms in the domain explicitly. The QC provides a framework where degrees of freedom are judiciously eliminated and force or energy calculations are approximated at areas of low interest, whereas exact discrete representation is used at the area of high interest such as crack tips or dislocation zones. The QC method also provides an adaptive framework for the fully resolved critical part to evolve during the simulation to balance between computational cost and error estimates.

The QC method finds the deformation field that minimizes the system potential energy while achieving the following three criteria \cite{Miller2002}:
\begin{enumerate}	
\item The number of degrees of freedom is reduced by limiting the degrees of freedoms to a small fraction of the nodes, called representative nodes (repnodes), but the full discrete description is retained in high interest regions, around crack tip for example. Hence, the computational cost of solving the system is significantly reduced. 
\item The computation of the total energy is accurately approximated without the need to explicitly compute the site energy of all the atoms. Hence, the computational cost associated with system assembly is significantly reduced.
\item The fully resolved critical regions can evolve with the deformation, during the simulation. Thus, the method achieves efficiency by using adaptivity to reduce the fully resolved domain as much as possible.
\end{enumerate}
\begin{figure}[ht]
\centering\includegraphics[width=0.75\linewidth]{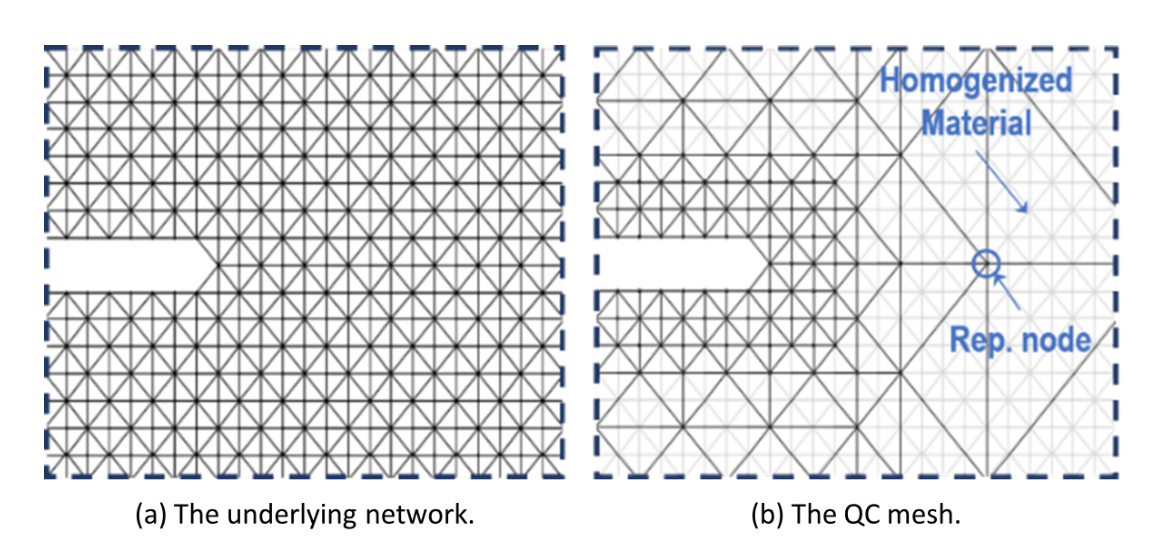}
\caption{A schematic for domain discretization for a notched sample with a lattice micro-structure using the Quasi-Continuum method: (a) the underlying network, and (b) the QC mesh with full network resolution around the crack, and homogenized material away from it. The degrees of freedom at the low interest area are limited to the repnodes.}
\label{fig:QC}
\end{figure}
The QC method applied to atomistic systems has been successfully used to investigate many localized phenomena, such as Nano-indentation, crack-tip deformation, deformation and fracture of grain boundaries, and dislocation interactions \cite{Tadmor1999NanoindentationPlasticity,Knap2001AnMethod,Miller_1998}. 

\textcolor{black}{While the QC method was originally introduced for atomic simulations, it may be extended to any discrete systems including networks or lattices. Beex et al. extended the QC method to regular discrete lattice networks with short-range nearest-neighbor interactions for both conservative \cite{Beex2011AModels}, and non-conservative lattice models such as dissipation, bond failure, and fiber sliding \cite{Beex2014ANetworks,Beex2014ASliding}, and used the method to model failure in electronic textiles \cite{Beex2015TheMethod}. The method was also extended to uniform linear elastic beam-like lattices experiencing planer bending \cite{Beex2015Higher-orderBending} and both in plane and out of plane deformation \cite{Beex2014Quasicontinuum-basedDeformation}. Rokos et al. used energy-based variational QC framework for modeling regular networks with linear lattices undergoing localized damage \cite{Rokos2016AMethods,Rokos2017AnDamage}. Mikes and Jirasek extended the quasi continuum method to irregular linear elastic lattices with axial interactions and small deformations \cite{Mikes2017QuasicontinuumLatticesb}. They applied the method to 
simulate nanotextile structures \cite{Mikes2016QuasicontinuumModel}, and model the microstructure based on the anisotropic microplane model \cite{Mikes2017QuasicontinuumModel}. Phlipot et al. \cite{Phlipot2018ALattices} used the method to model linear elastic periodic beam lattices with co-rotational framework.}

\textcolor{black}{In addition to lattice structures, the QC methodology has been extended for simulation of granular media where the discrete element method is used at areas of high interest while homogenization is utilized elsewhere for computational efficiency \cite{Wellmann2012AMaterials,Yue2018HybridMedia}. The flexibility of the framework makes it a suitable candidate for applications to irregular polymer networks with both material and geometric nonlinearities, which is the focus of this study.} 

\section{Quasi-Continuum for irregular polymer networks:}

In this section, we discuss the problem setup and the finite element formulation for both the discrete part and the homogenized part of the network model. For the discrete part, we use a nonlinear 1D finite element with finite deformation and rotation. For the homogenized part, we use 2D nonlinear triangular finite element with finite deformations. The material properties for the 2D elements are derived by applying homogenization rule based on the micro-macro energy consistency condition that conserves the variation of local work between the macro (2D homogenized elements) and micro (underlying network) scales. 
\subsection{Finite Element Formulation: 2D homogenized elements}
The strong form for the boundary value problem in the absence of inertia effects is given by:

\begin{equation}
\frac{\partial P_{ik}}{\partial X_k}+\rho_0 b_i = 0
\end{equation}
Where ${\bf P}$ is the $1^{st}$ Piola-Kirchhoff stress tensor, ${\bf X}$ is the material point coordinate in the reference configuration,  ${\bf b}$ is the body force, and $\rho_0$ is the material density at the reference configuration.

The boundary conditions are given by:

\begin{equation}
	P_{ij}N_j=\bar{t}_i\;\;\; \text{on} \;\;\; \Gamma^o_{t_i}\;\;\; \text{and} \;\;\;
	u_{i}=\bar{u}_i\;\;\; \text{on} \;\;\; \Gamma^o_{u_i}
\end{equation}
Where
	\[\Gamma^o_{t_i} \cap \Gamma^o_{u_i}=\phi \;,\;\Gamma^o_{t_i} \cup \Gamma^o_{u_i}=\Gamma^o\]
Where $\bar{t}_i$ is the applied traction on boundary $\Gamma^o_{t_i}$, $\bf{N}$ is the unit vector normal to the boundary surface in the reference configuration, $\bar{u}_i$ is the prescribed displacement on boundary $\Gamma^o_{u_i}$, and $\Gamma^o$ is the boundary surface.

The weak form in a total Lagrangian formulation may be written as:

\begin{equation}
	\int_{{\Omega_o}} \delta{F_{ik}} P_{ik} d\Omega_o  = \int_{{\Omega_o}} \delta{u_{i}} \rho_0 {b}_i d\Omega_o + \int_{{\Gamma^o_{t_i}}} \delta{u_{i}} \bar{t}_i \Gamma^o
\end{equation}
Where $\mathbf{F}$ is the deformation gradient, and $\delta \mathbf{u} \in \mathbf{u}_o$ where $\mathbf{u}_o$ the space of kinematically admissible displacements with the requirement that the displacements vanish on displacement boundaries. 
The finite element approximation is derived in terms of the objective $2^{nd}$ Piola-Kirchhoff stress tensor $\mathbf{S}$. After lineraization and making use of Voigt notation, the internal force vector and material tangent matrix are given by:
\begin{equation}
	\mathbf{f}^{int}_I=\int_{{\Omega_o}} \mathbf{B}_{oI}^T {\mathbf{S}} d\Omega_o \quad \quad \text{and} \quad \quad \mathbf{K}_{IJ}=\mathbf{K}_{IJ}^{mat}+\mathbf{K}_{IJ}^{geo}
\end{equation}
Where
\begin{equation}
	\mathbf{K}_{IJ}^{mat}=\int_{{\Omega_o}} \mathbf{B}_{oI}^T [\mathbf{C}^{SE}] \mathbf{B}_{oJ} d \Omega_o \quad \quad \text{and} \quad \quad \mathbf{K}_{IJ}^{geo}=\mathbf{I} \int_{{\Omega_o}} \mathbf{B}_{oI}^T \mathbf{S} \mathbf{B}_{oJ} d \Omega_o
\end{equation}
Where $\mathbf{B}_o$ is the strain-displacement matrix. The goal of the homogenization is to derive the macroscopic stress $\mathbf{S}$ and material tangent  $\mathbf{C}^{SE}$ tensors that map the micro-scale behavior with sufficient accuracy. These quantities are then provided to the nonlinear finite element framework for solving the system.

\subsection{Homogenization:}

\textcolor{black}{We replace the network links, away from the high interest zones, by continuum elements with material properties obtained by homogenization of the discrete network. This results in a significant number of network links being removed from the assembly procedure, leading to significant computational savings. This approach was pioneered by Mikes and Jirasek \cite{Mikes2017QuasicontinuumLatticesb} to model irregular linear elastic lattices with axial interactions and small deformations. In this study, we extend the formulation to nonlinear links with finite deformations.} 
\subsubsection{Micro-macro energy consistency condition:}
We establish the micro-macro scale transition relation based on the Hill-Mandel condition \cite{Hill1963ElasticPrinciples,edselc.2-52.0-002219985519850101}. This condition requires the volume average of the variation of work on the micro level to be equal to the variation of local work on the macroscale: 
\begin{equation}
    \delta W_{Macro}=\delta W_{micro}
\end{equation}
In terms of macro deformation gradient tensor and the first Piola-Kirchhoff stress tensor, the condition reads \cite{Geers2017HomogenizationProblems}:
\begin{equation}\label{eq:P1}
    \mathbf{P}_M:\delta \mathbf{F}_M = \frac{1}{V_o} \sum_{p=1}^{N_p} \left(\vec{f}_p.\delta \vec{l}_p\right)
\end{equation}
Where $\vec{f}_p$ and $\vec{l}_p$ are the force and the length of each link in the current configuration, $N_p$ is the number of links.
Using $\delta \vec{l}_p =  \vec{L}_p \delta \mathbf{F}^T $, where $\vec{L}_p$ is the link length in the original configuration, Eq.\ref{eq:P1} becomes:
\begin{equation}
    \mathbf{P}_M:\delta \mathbf{F}_M = \frac{1}{V_o} \sum_{p=1}^{N_p} \left(\vec{f}_p\otimes \vec{L}_p\right):\delta \mathbf{F}_m
\end{equation}
Where the subscript $M$ denotes the macro level tensors, and $m$ denotes the micro level tensors. Using Voigt assumption (i.e $\mathbf{F}_m=\mathbf{F}_M$). The homogenized $1^{st}$ Piola-Kirchhoff stress is given by:
\begin{equation} \label{eq:PM}
\mathbf{P}_M = \frac{1}{V_o} \sum_{p=1}^{N_p} \left(\vec{f}_p\otimes \vec{L}_p\right)
\end{equation}

To calculate the tangent tensor, we start by taking the variation of $\mathbf{P}_M$:

\begin{equation}
    \delta \mathbf{P}_M = \frac{1}{V_o} \sum_{p=1}^{N_p} \left(\frac{\partial \vec{f}_p}{\partial \vec{l}_p}. \delta \vec{l}_p \right)\otimes \vec{L}_p
\end{equation}
Substituting $\delta \vec{l}_p =  \delta \mathbf{F} \vec{L}_p$ and rearranging:

\begin{equation}
    \delta \mathbf{P}_M = \frac{1}{V_o} \sum_{p=1}^{N_p} \left(\vec{L}_p \otimes \frac{\partial \vec{f}_p}{\partial \vec{l}_p} \otimes \vec{L}_p\right)^{LT}:\delta \mathbf{F}_m
\end{equation}
Where the subscript LT denotes the transpose of the left two indices. Again, using Voigt assumption, it follows that:

\begin{equation}
    \mathbf{C}^{PF}=\frac{1}{V_o} \sum_{p=1}^{N_p} \left(\vec{L}_p \otimes \frac{\partial \vec{f}_p}{\partial \vec{l}_p} \otimes \vec{L}_p\right)^{LT}
    \label{eq:CPF}
\end{equation}
Where $\mathbf{C}^{PF}$ is the homogenized $4^{th}$ order first elasticity tensor. It must be noted that the homogenized stress tensor is a two-point tensor (i.e not symmetric), and the elasticity tensor has only major symmetry.

It is more convenient to use the objective symmetric $2^{nd}$ Piola-Kirchhoff Stress tensor and the corresponding elasticity tensor. By definition, the $2^{nd}$ Piola-Kirchhoff Stress tensor is given by:
\begin{equation}
    \mathbf{S}_M=\mathbf{F}_M^{-1}\mathbf{P}_M
\end{equation}
Using indicial notation and substituting for $\mathbf{P}_M$ from Eq.\ref{eq:PM}
\begin{equation}
    S^M_{ij}={F^M_{ik}}^{-1}\frac{1}{V_o} \sum_{p=1}^{N_p} \left({f}^p_k  {L}^p_j\right)
\end{equation}
After using Voigt assumption:
\begin{equation}
    S^M_{ij}=\frac{1}{V_o} \sum_{p=1}^{N_p} \left({F^p_{ik}}^{-1}{f}^p_k  {L}^p_j\right)
\end{equation}
We note that the quantity ${F^p_{ik}}^{-1}{f}^p_k$ is the (fictitious) force vector corresponding to the PK2 stress.

To derive the elasticity tensor relating the PK2 tensor and the Green strain, we start by varying the relation between PK1 and PK2:

\begin{equation}
    \delta {S}^M_{ij}=\underbrace{{F^M_{im}}^{-1} \delta {P}^M_{mj}}_{Term 1} +\underbrace{\delta {F^M_{iu}}^{-1} {P}^M_{uj}}_{Term 2}
\end{equation} 
After substituting from Eq.\ref{eq:CPF} and using Voigt assumption, the first term may then be written as: (we drop the subscript p for convenience)

\begin{equation}
    {F^M_{im}}^{-1} \delta {P}^M_{mj} = \frac{1}{V_o} \sum_{p=1}^{N_p} \left(F_{im}^{-1} {L}_j \frac{\partial {f}_m}{\partial {l}_n} {L}_l\right):\delta F_{nl}
\end{equation}
Simplifying and using $\delta E_{kl}=F_{nk} \delta{F}_{nl}$, $\delta E_{kl}=\delta E_{lk}$:
\begin{equation}
    {F^M_{im}}^{-1} \delta {P}^M_{mj} = \frac{1}{V_o} \sum_{p=1}^{N_p}   \left({L}_j F_{im}^{-1} \frac{\partial {f}_m}{\partial {l}_n} F_{kn}^{-1} {L}_l\right):\delta E_{kl}
\end{equation}
The second term is written as:
\begin{equation}
    \delta {F^M_{iu}}^{-1} {P}^M_{uj}= \frac{\partial F_{iu}^{-1}}{\partial F_{nk}} \delta F_{nk} {P}_{uj}
\end{equation}
After simplification and some indices manipulation:
\begin{equation}
    \delta {F^M_{iu}}^{-1} {P}^M_{uj}=-F_{in}^{-1}  F_{ku}^{-1} {P}_{uj} : \delta F_{nk} 
\end{equation}

\begin{equation}
    \delta {F^M_{iu}}^{-1} {P}^M_{uj}=-F_{in}^{-1} S_{kj}:\delta F_{nk}  
\end{equation}

\begin{equation}
    \delta {F^M_{iu}}^{-1} {P}^M_{uj}= -F_{in}^{-1} F_{ln}^{-1}  S_{kj} :\delta E_{lk}  
\end{equation}
Hence:

\begin{equation}
    {C}^{SE}_{ijkl}=\frac{1}{V_o} \sum_{p=1}^{N_p} \left({L}_j F_{im}^{-1} \frac{\partial {f}_m}{\partial {l}_n} F_{kn}^{-1} {L}_l\right) - F_{in}^{-1} F_{ln}^{-1}  S_{jk}
\end{equation}
Where $\mathbf{C}^{SE}$ is the homogenized $4^{th}$ order second elasticity tensor. 
\subsection{Assembling the tensors:}
Since the links in the network have axial deformations only, the deformation gradient for each link is written as:
\begin{equation}
    F_{ij}=\lambda n_i N_j
\end{equation}
Where $\lambda = l/L$ is the stretch, $n_i$ is the unit vector in the current configuration, and $N_i$ is the unit vector in the original configuration. Similarly:
\begin{equation}
    F_{ij}^{-1}=\frac{1}{\lambda} N_i n_j
\end{equation}
The PK1 stress tensor is then written as
\begin{equation}
    P_{ij}=\frac{1}{V_o} \sum_{p=1}^{N_p} \left( f  L \right)n_i N_j
\end{equation}
And the PK2 stress tensor:
\begin{equation}
    S_{ij}=\frac{1}{V_o} \sum_{p=1}^{N_p} \left(\frac{1}{\lambda} f  L \right)N_i N_j
    \label{eq:S}
\end{equation}
Whereas the material tensors are given by
\begin{equation}
    {C}^{PF}_{ijkl}=\frac{1}{V_o} \sum_{p=1}^{N_p} \left( k L^2  \right)n_i N_j n_k N_l
\end{equation}
\begin{equation}
    {C}^{SE}_{ijkl}=\frac{1}{V_o} \sum_{p=1}^{N_p} \left(\frac{1}{\lambda^2} k L^2  - \frac{1}{\lambda^3} f  L \right)N_i N_j N_k N_l
    \label{eq:Cse}
\end{equation}
Here $k=\partial f/\partial l$ for the link, $f$, and $L$ are the magnitudes of the link force and reference length respectively. The stress tensor $\mathbf{S}$ is symmetric, and the elasticity tensor $\mathbf{C}^{SE}$ has both major and minor symmetries.

\subsection{Polymer chain constitutive law:}
In this study, we model each link in the polymer network using a nonlinear elastic force elongation relation give by the worm like chain model. The force-elongation relation is given by \cite{Marko1995StretchingDNA}:
\begin{equation}
\label{eq:pcf}
f=\frac{k_B T}{b}\left[\frac{1}{4}\left(1-\frac{x}{L_c}\right)^{-2}-\frac{1}{4}+\frac{x}{L_c}\right]
\end{equation}
Where $f$ is the force, $x$ is the end-to-end distance, $b$ is the persistence length, $k_B$ is Boltzmann' constant, $T$ is the temperature, and $L_c$ is the contour length of the polymer chain. We assume the network is force balanced at the reference configuration. The chain stiffness is given by the slope of the tangent of the force elongation curve and it takes the form

\begin{equation}
\label{eq:pck}
k=df/dx=\frac{k_B T}{b}\left[\frac{1}{2L_c}\left(1-\frac{x}{L_c}\right)^{-3}+\frac{1}{L_c}\right]
\end{equation}

These expressions are used in Eq.\ref{eq:S} and Eq.\ref{eq:Cse} to compute the homogenized stress and elasticity tensor. As the end-to-end distance $x$ approaches the contour length $L_c$, the polymer chain response becomes highly nonlinear as both the force and the stiffness values go to infinity. This signals a limitation of this constitutive description which may be circumvented by accounting for strain energy of the chain in addition to its entropic energy \cite{Marko1998DNADynamics}. Such correction will be investigated further in future work. In this initial study, such extreme limit is not probed as we adopt a maximum stretch failure criterion for the links. That is, we assume the link would fail if its stretch relative to the available contour length $\lambda_p = l/L_c$ exceeds a threshold value that is predefined. Other failure criterion based on transition state theory for bond breakage \cite{Lieou2013SacrificialBone} may also be used.

\section{Numerical Implementation:}\label{sec:NumImp}
To couple the micro- and macro-scales, the network nodal displacements $u_n$ are related to the representative nodes (repnodes) displacement $u_R$, see Fig.\ref{fig:QCvsD}-c, through the interpolation functions:
\begin{equation}
    u_n=\sum_{i=1}^{N_{rep}} \phi_R u_R 
    \label{eq:intarpolation}
\end{equation}
where $\phi_R$ is the interpolation function associated with repnode $R$ and $N_{rep}$ is the number of repnodes. Using the compact support of the finite element shape functions, the displacement of each node is determined from the sum over the three vertices of the triangle containing this node. In this initial study, we use linear shape functions which turns out to perform satisfactorily as will be discussed shortly.

After evaluating the network nodal displacements, the stress and material tangent tensors are assembled as follows:

\begin{algorithm}[H]
\SetAlgoLined
     Calculate the network nodal displacements from the last QC solution through interpolation functions given in Eq.\ref{eq:intarpolation}\;
     \For{each QC element $\Omega_i$}{
        \For{each link inside element $\Omega_i$}{
        calculate link force $f$ from Eq.\ref{eq:pcf}\;
        calculate link stiffness $k$ from Eq.\ref{eq:pck}\;
        calculate link stretch $\lambda=l/L$\;
        update $\mathbf{S}$ using Eq.\ref{eq:S}\;
        update $\mathbf{C}^{SE}$ using Eq.\ref{eq:pcf}\;
        }}
     \caption{Homogenization of QC elements}\label{algo:hom.}
\end{algorithm}

\textcolor{black}{Since we use the total Lagrangian formulation, the reference configuration quantities, such as the reference length $L$, and the reference unit vector $N$, are calculated at the system initialization and are stored instead of calculating them each time step. This leads to further computational savings}. \textcolor{black}{As discussed in Section \ref{sec:overview}, QC methodologies normally use two reduction steps: limiting the degrees of freedom to a small fraction of the nodes and sampling of the lattice interactions for efficient energy summation \cite{Beex2014ANetworks,Rokos2017AnDamage}. However, in the current work, due to the network irregularity, we account for the contribution of all links in each continuum element in the homogenization procedure \cite{Mikes2017QuasicontinuumLatticesb}. We discuss possible alternative approaches in Section \ref{sec:discussion}.} 

\subsection{Automatic mesh adaptivity:}
Automatic Mesh Adaptivity is a critical ingredient for the efficient implementation of the QC method as it enables the QC mesh to evolve dynamically based on the crack propagation path. In our proposed approach, we mark a 2D finite element $\Omega_i$ for refinement if it meets 1 of the following criteria:
\begin{enumerate}
    \item Link stretch criterion:
    \begin{equation}
        \lambda_p > \lambda_{th} \;\;\; p \in \Omega_i
    \end{equation}
    Where $\lambda_p = l/L_c$ is the stretch relative to the polymer chain contour length of any network link inside the QC element $\Omega_i$, and $\lambda_{th}$ is a threshold value. The threshold value is chosen as a fraction of the failure stretch.
    \item Deformation gradient error criterion \cite{Miller2002}:
    \begin{equation}
         \varepsilon_i=\bigg(\frac{1}{V_{\Omega_i}} \int_{\Omega_i} (\bar{\mathbf{F}} -\mathbf{F}_i):(\bar{\mathbf{F}} -\mathbf{F}_i) d {V}\bigg)^\frac{1}{2} > \varepsilon_{th}
    \end{equation}
    
    Where $\varepsilon_i$ is a scalar measure that quantifies the error introduced into the solution by the current density of representative nodes, $V_{\Omega_i}$ is the volume of element $\Omega_i$, $\mathbf{F}_i$ is the QC solution for the deformation gradient in element $\Omega_i$, $\bar{\mathbf{F}}$ is the $L_2$-projection of the QC solution for $\mathbf{F}$ given by $\bar{\mathbf{F}} = \phi \mathbf{F}_{avg}$ where $\phi$ is the shape function array, and $\mathbf{F}_{avg}$ is computed by averaging the deformation gradient in each element sharing a given repnode, and $\varepsilon_{th}$ is a specified threshold value.
\end{enumerate}

\begin{figure}[H]
\centering\includegraphics[width=1\linewidth]{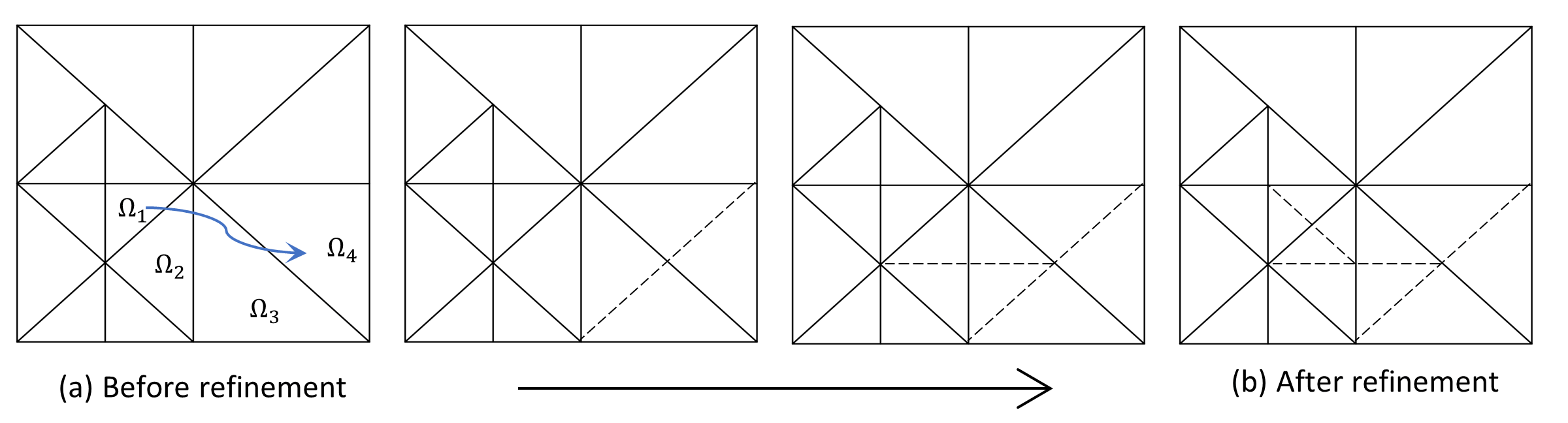}
\caption{Mesh refinement steps: (a) the initial mesh where $\Omega_1$ is marked for refinement, and (b) the final mesh with the newly added edges shown as dashed lines. The LEPP for $\Omega_1$ is tracked and triangles are bisected in backward manner till $\Omega_1$ is refined.}
\label{fig:adap}
\end{figure}

We start by constructing right-triangulated initial mesh to reduce summation errors as per \cite{Rokos2017EXtendedPropagation}. Each QC mesh point is then moved to the nearest network node. All elements marked for refinement are added to a set $E_{ref}$. For each element inside $E_{ref}$. We use the standard Rivera algorithm \cite{Rivara1997NewTriangulations}, which conserves non-degeneracy, conformity, and smoothness. This algorithm tracks the longest-edge propagation path (LEPP) associated with a triangular element $\Omega_i$ in a backward manner.
The LEPP of $\Omega_i$ in a conforming triangular mesh is defined as an ordered
list of triangles $\Omega_i$, $\Omega_{i+1}$,..., $\Omega_n$ such that $\Omega_{i+1}$ is a neighbour by the longest edge of $\Omega_i$ for each  $i = 1,2, ..., n$, the refinement scheme is shown in Fig.\ref{fig:adap}. The algorithm, which is summarized in Algorithm 2, is then used for each $\Omega_i \in E_{ref}$ and repeated till $\Omega_i$ is fully resolved.

\begin{algorithm}[H]
\SetAlgoLined
 \For{$\Omega_i \in E_{ref}$}{
 \While{$\Omega_i$ isn't bisected}{
  Find the LEPP of $\Omega_i$\;
   \eIf{The longest edge of $\Omega_n$, the last triangle in LEPP of $\Omega_i$, is on $\Gamma^o$}{
   Bisect $\Omega_n$\;
   }{
   Bisect the last two triangles in LEPP of $\Omega_i$,  ($\Omega_n,\Omega_{n-1}$)\;}}
  Move each rep node to the nearest network node\;
  \If{$\Omega_i$ is fully resolved}{
   Transform from QC element into discrete elements\;
   }
   }
 \caption{Mesh adaptivity Algorithm} \label{alg:ref}
\end{algorithm}

After mesh refinement, each representative node is moved to the nearest network node. \textcolor{black}{All fully refined QC elements (i.e. elements that contain only 3 network nodes) are transformed from 2D triangular elements to 1D links consistent with the underlying network structure. This step also prevents ill-conditioning of the stiffness matrix of the QC system \cite{Beex2014Quasicontinuum-basedDeformation}.}

\subsection{Time adaptivity:}
Time adaptivity is required for computational efficiency by allowing larger time steps, when possible, without compromising accuracy. In addition, time adaptivity is also critical for the accurate modeling of fracture in high interest zones where small time steps are needed. Using a constant time step may be overly restrictive if a small time step is used throughout and may lead to inaccurate results if a large time step is perpetually maintained. Time adaptivity balances computational cost and error. \textcolor{black}{The simulation starts with a stable large time step, that accurately and efficiently compute the nonlinear elastic deformation. The algorithm has two time refinement criteria which are usually triggered when the discrete elements approach failure. The first one is to ensure convergence where the time step is reduced, if needed, to reach the stable time step required for the solution convergence. The second criterion is to ensure accuracy in predicting the crack propagation as large time steps may lead to overestimating the number of failed links. Once failure initiates, the time step is reduced progressively such that further reduction in the step size won't affect the number of elements failing at the current time step. In the current study, the time step is the same for both the homogenized and fully discrete domains. We discuss possible alternative approaches in Section \ref{sec:discussion}.}

\subsection{Nonlinear finite element framework:}
The algorithm used for adaptive modeling of fracture in polymer networks is given in Algorithm \ref{alg:FEM}. The homogenization is performed on the fly to provide the stress and the material tangent tensor each time step.

\begin{algorithm}[H]
\SetAlgoLined
     Initialize the system, construct the initial mesh with required information\;
     \For{step = 1,2,......,n}{
      Calculate the homogenized material consistent tangent and stress tensor using Algorithm (\ref{algo:hom.})\;
      Apply the step incremental boundary conditions at step n\;
      Assemble the system internal force vector and tangent matrix and solve for the displacement\;
      For each 2D element check the mesh adaptivity conditions, add all elements marked for refinement to a set $E_{ref}$\;
      \eIf{$E_{ref}$ is not empty}{
       Refine the current mesh as per Algorithm \ref{alg:ref}, update system information\;
       Repeat the current time step to balance the new mesh system\;
       }{
       Store the relevant outputs for the current time step\;
       Proceed to the next time step\;
      }
     }
     \caption{Adaptive QC based finite element framework}
     \label{alg:FEM}
\end{algorithm}

\section{Verification}\label{sec:ver}
For verification of the proposed homogenization technique, we run multiple tests for both pristine and cracked samples. In each case, we verify the QC results by comparing it to the results of the fully discrete model and quantify the error in terms of the measured reaction normalized force at different levels of stretch. We define the normalized force as $fb/k_Bt$ (Eq.\ref{eq:pcf}).

\subsection{Uniform deformation gradient tests:}
\begin{figure}[H]
\centering\includegraphics[width=0.8\linewidth]{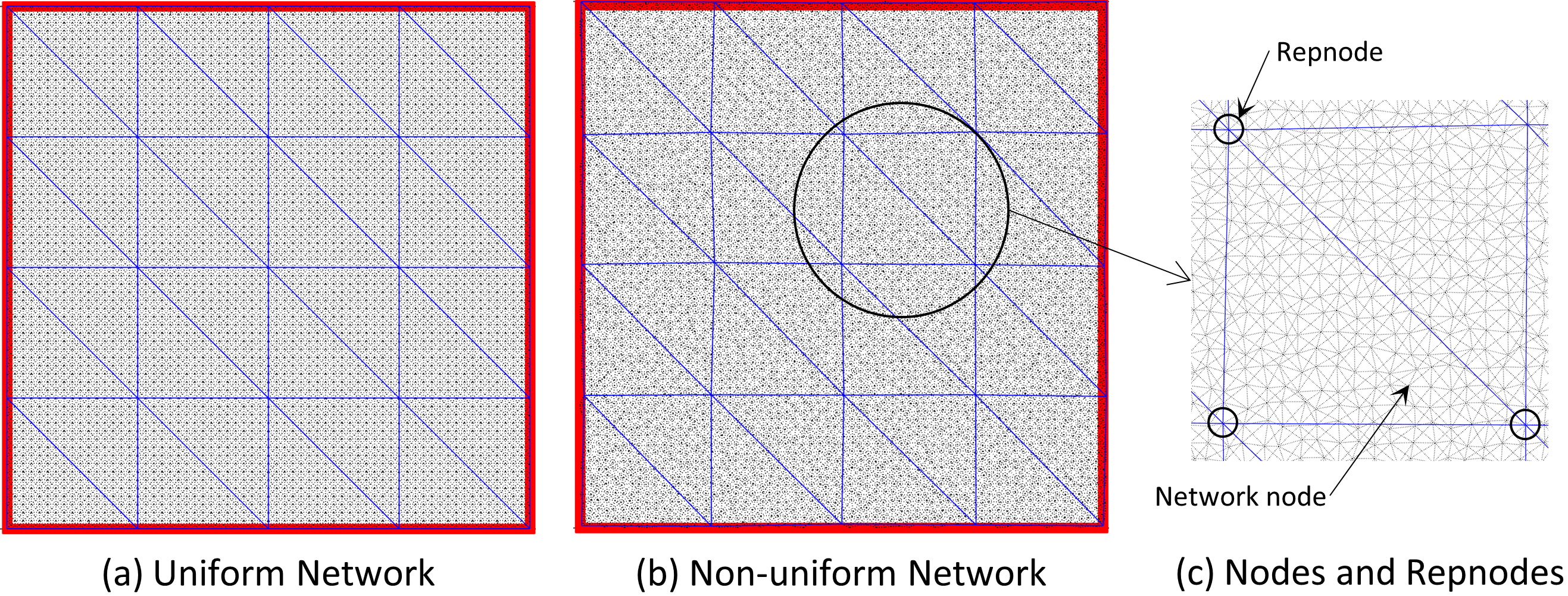}
\caption{The pristine sample used for verification: (a) uniform network (in black) and the corresponding QC mesh (in blue), (b) non uniform network where the nodal location is shifted randomly from the uniform case, the maximum shift is controlled by the disorder parameter $\mu_g$, and (c) a QC element showing the network nodes inside this element and the repnodes used for interpolation.}
\label{fig:QCvsD}
\end{figure}

 For the first test, we consider a regular network geometry shown in Fig.\ref{fig:QCvsD}-a with uniform material properties. All the network links in the sample are homogenized using the procedure outlined in Section \ref{sec:NumImp} . We impose an affine deformation gradient, considering both uni-axial and bi-axial tension loading cases, up to a stretch value of 6.  The QC results  match exactly the fully discrete model in terms of the reaction force and the total strain energy values. This verifies that the homogenization scheme recovers the exact response if the microscale deformation is affine and the network structure is regular.
 
\begin{figure}[ht]
\centering\includegraphics[width=1.0\linewidth]{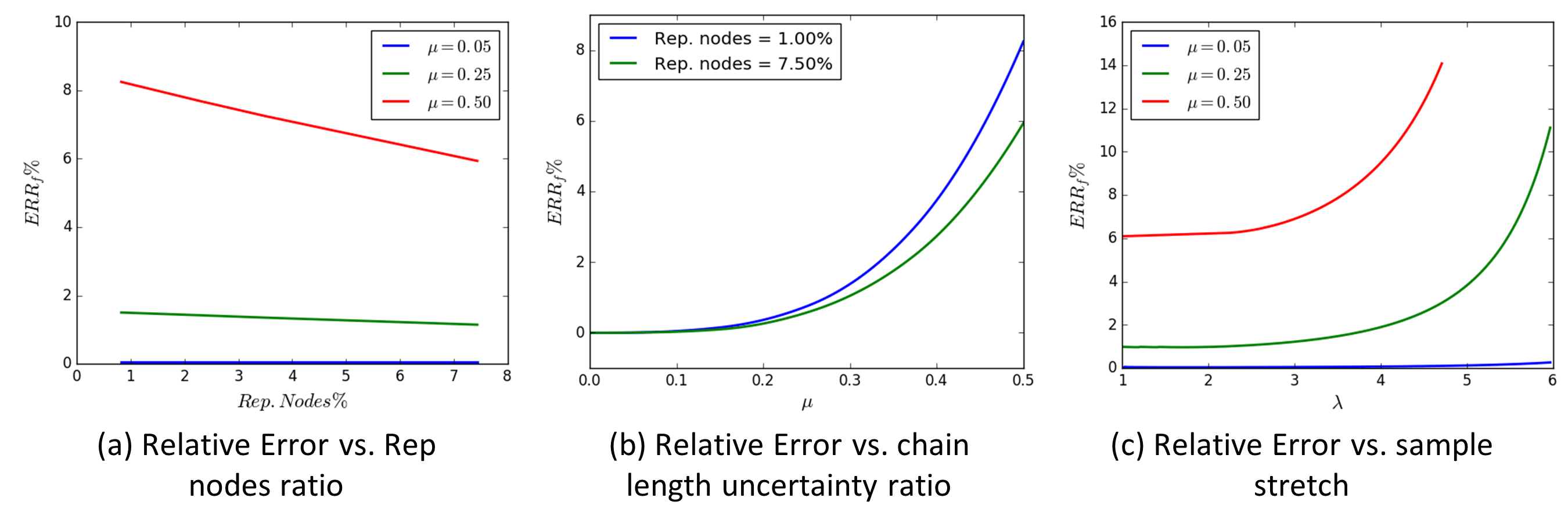}
\caption{Homogenization error of the QC method applied to disordered networks: (a) the relative error in the force at $\lambda = 1$ vs. the ratio of the repnodes to the network nodes for 3 different disorder parameter $\mu = 0.05 \text{, } 0.25 \text{, and } 0.5$, (b) the relative error in the force at $\lambda = 1$ vs. $\mu$ for two values of repnodes ratio, and (c) the evolution of the error vs. stretch for $\mu = 0.05 \text{, } 0.25 \text{, and } 0.5$ and repnodes ratio = 7.5\%.}
\label{fig:VerRand}
\end{figure}

Next, to study the effect of disorder in the geometric and material properties on the performance of the QC method, we simulate the response of a non uniform network, as shown in Fig.\ref{fig:QCvsD}-b and the zoomed in view in Fig.\ref{fig:QCvsD}-c, under uni-axial tension. We introduce non-uniformity in two ways. First, we shift the position of the network nodes randomly with respect to a reference perfectly ordered lattice. The maximum shift values in $x$ and $y$ directions are $\mu_g L_x$ and $\mu_g L_y$, respectively, where $\mu_g$ is the geometric nonconformity parameter. Second, we draw the polymer chain contour length $L_c$  randomly from a uniform distribution ranging between $[(1-\mu)L_c,(1+\mu)L_c]$, where $\mu$ is the contour length nonconformity parameter. For each non-uniformity parameter value, ten randomly generated meshes are simulated and the average results are plotted in Fig.\ref{fig:VerRand}. We define the relative error in the force as:
\begin{equation}
    ERR_f = \frac{|\sum_k R_k^{D}-\sum_k R_k^{QC}|}{\sum_k R_k^{D}}
\end{equation}
Where $\sum_k R_k$ is the sum of reactions in the loading direction, $D$ refers to the corresponding fully discrete simulation and $QC$ refers to the QC simulations containing homogenized continuum elements only.

Fig.\ref{fig:VerRand}-a shows that as the disorder increases the error percentage from the homogenization scheme increases. However, this error decreases as the fraction of repnodes in the sample increases. This is further confirmed in Fig.\ref{fig:VerRand}-b which shows that variation of error as a function of variability in the chain lengths at two values of repnodes percentages. At low values of disorder, $\mu$ less than 0.2, both the homogenized and discrete models give identical results (the error is almost zero). As the disorder increases, the error increases. However, it remains less than 8 \%  and it decreases with mesh refinement. Fig.\ref{fig:VerRand}-c shows that the error from the homogenization depends on the stretch level. For low values of disorder, the error is negligible even at high stretch values (up to 6). As the disorder increases, the error is relatively small (few percents) at low stretch values but it rapidly increases at high stretch values reaching as high as 15 \%  for $\mu$=0.5 at stretch value of 5. While this exercise attests to the accuracy of the homogenization scheme at low stretch values, it suggests that mesh refinement is crucial for retaining accuracy at high level of stretch. This will be accounted for automatically in the QC application by turning on dynamic adaptivity and enforcing a refinement criterion based on the local stretch level as explained in Section \ref{sec:NumImp}.

\subsection{Non-uniform deformation gradients-- Effect of refinement:}
\begin{figure}[H]
\centering\includegraphics[width=0.90\linewidth]{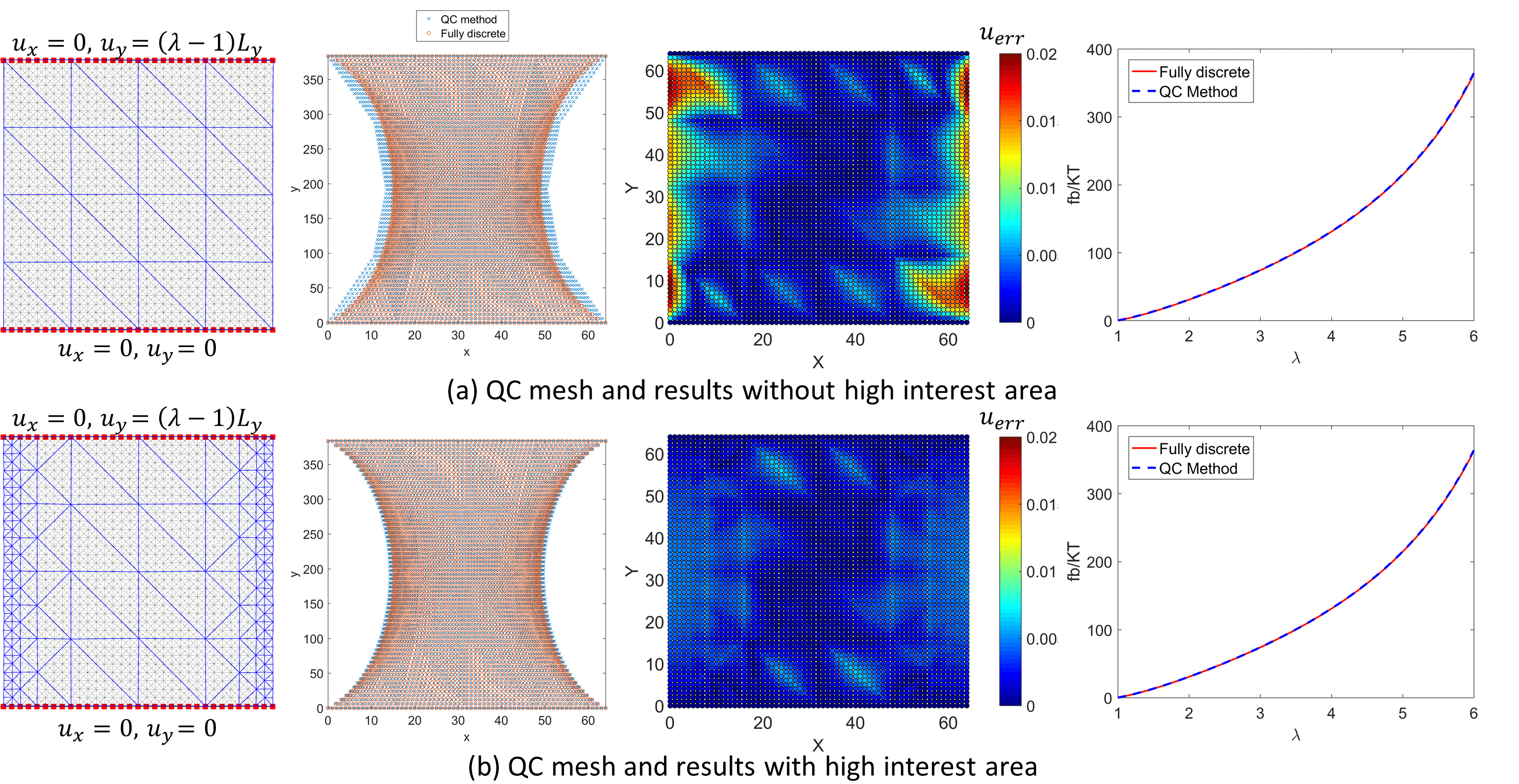}
\caption{The effect of refinement on the QC method applied to pristine networks under tensile loading: The network and the QC mesh, deformed shape, relative error in the nodal displacement, and force vs. stretch from both discrete and QC results for (a) QC mesh where no high interest area (i.e. no discrete links) is defined, and (b) QC mesh where the elements adjacent to the free vertical edges are refined to the discrete representation.}
\label{fig:Ver1}
\end{figure}

Here, we run two tests where the deformation gradient is not uniform over the domain.
In the first test, we model a pristine sample with dimensions 64x64 where the bottom edge is restrained from movement in both directions and the top edge is clamped and is pulled upward. The results shown in Fig.\ref{fig:Ver1} using only continuum elements, homogenized through the QC procedure, show that although the force vs. stretch shows good agreement with the full discrete model, up to a stretch value of 6, there is a discrepancy in the deformed shape specially near the edges. The error distribution plot also shows that the nodal displacement error is highest in the edge elements, closest to the free lateral boundaries, where the network nodal displacements are interpolated from the edge nodes of the continuum elements. To address this problem, we designate these areas as high interest areas and model the network links in these regions explicitly. This leads to eliminating the concentrated errors near the edges. Although the repnodes ratio increased from 0.5\% to around 5\%, the deformed shapes and the force vs. stretch curves using both QC and fully discrete match better in this case. Furthermore, the error in the displacement distribution falls almost an order of magnitude throughout the domain. In practical applications of the QC method, the discovery of high interest areas and the change in the model resolution is done automatically using the adaptive mesh refinement algorithm.

\begin{figure}[H]
\centering\includegraphics[width=0.90\linewidth]{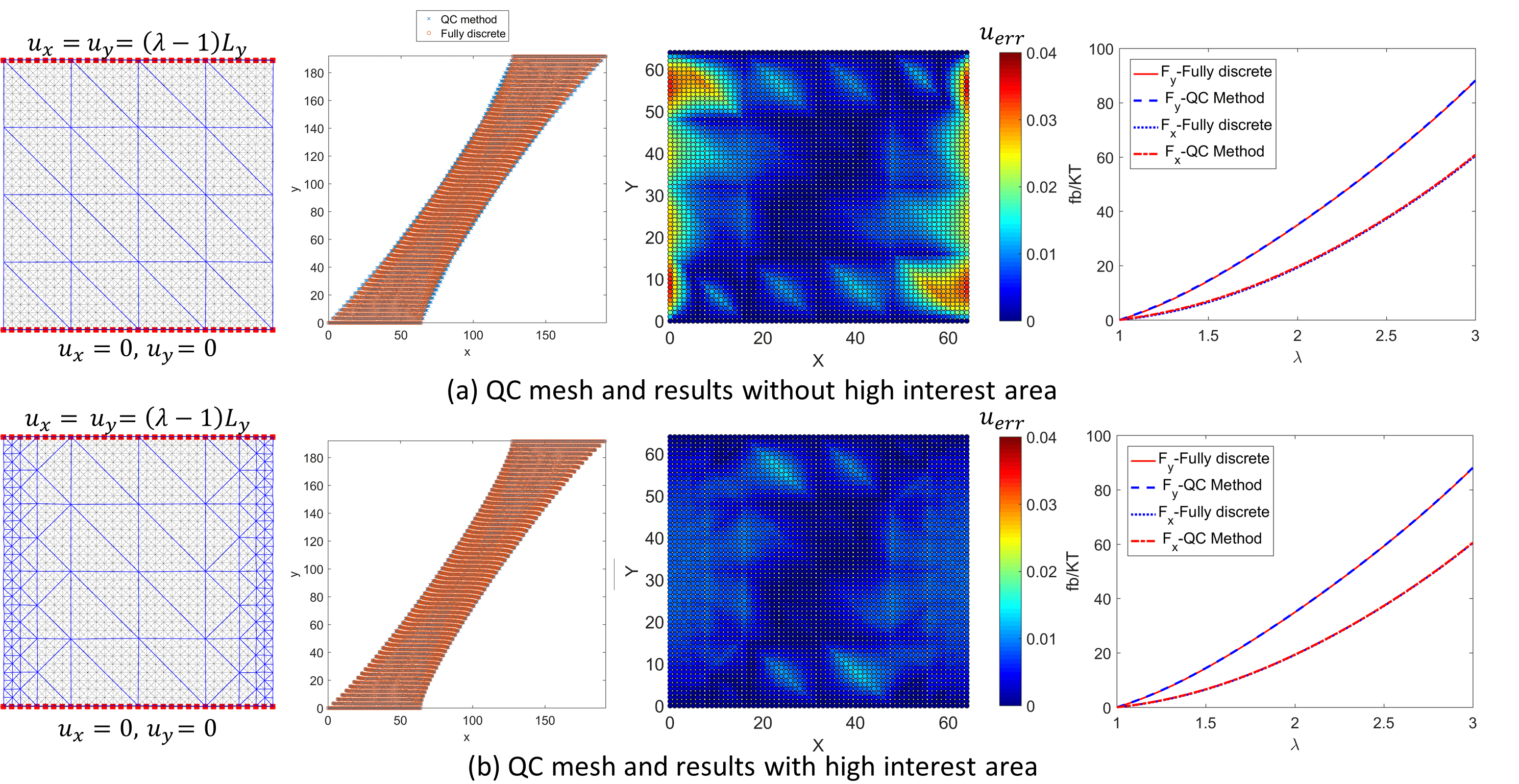}
\caption{The effect of refinement on the QC method applied to pristine networks under combined tensile and shear loading: The network and the QC mesh, deformed shape, relative error in the nodal displacement and force vs. stretch from both discrete and QC results for (a) QC mesh where no high interest area (i.e. no discrete links) is defined, and (b) QC mesh where the elements adjacent to the free edges are refined to the discrete representation.}
\label{fig:Ver2}
\end{figure}

In the second test, we use the same geometry and boundary conditions as in the first one, but the top edge now moves both upward and to the right such that the sample is subjected to both tension and shear. In this case, mesh refinement reduces the error in nodal displacement and leads to a better agreement in the force vs. stretch response between the QC and the full discrete models. The results shown in Fig.\ref{fig:Ver1} and Fig.\ref{fig:Ver2} illustrate the capability of the the QC method in bridging different scales, and also highlights the critical role of mesh adaptivity in reducing the errors specially in areas where the deformation gradient is not uniform. Furthermore, since the results in Fig.\ref{fig:Ver1}, and Fig.\ref{fig:Ver2} suggest that the error in the displacement is usually high near the free surfaces, we have chosen to always use an initial QC mesh that is refined around the sample boundaries in the remaining cases presented in this paper.

\subsection{Fracture test:}
\begin{figure}[H]
\centering\includegraphics[width=0.75\linewidth]{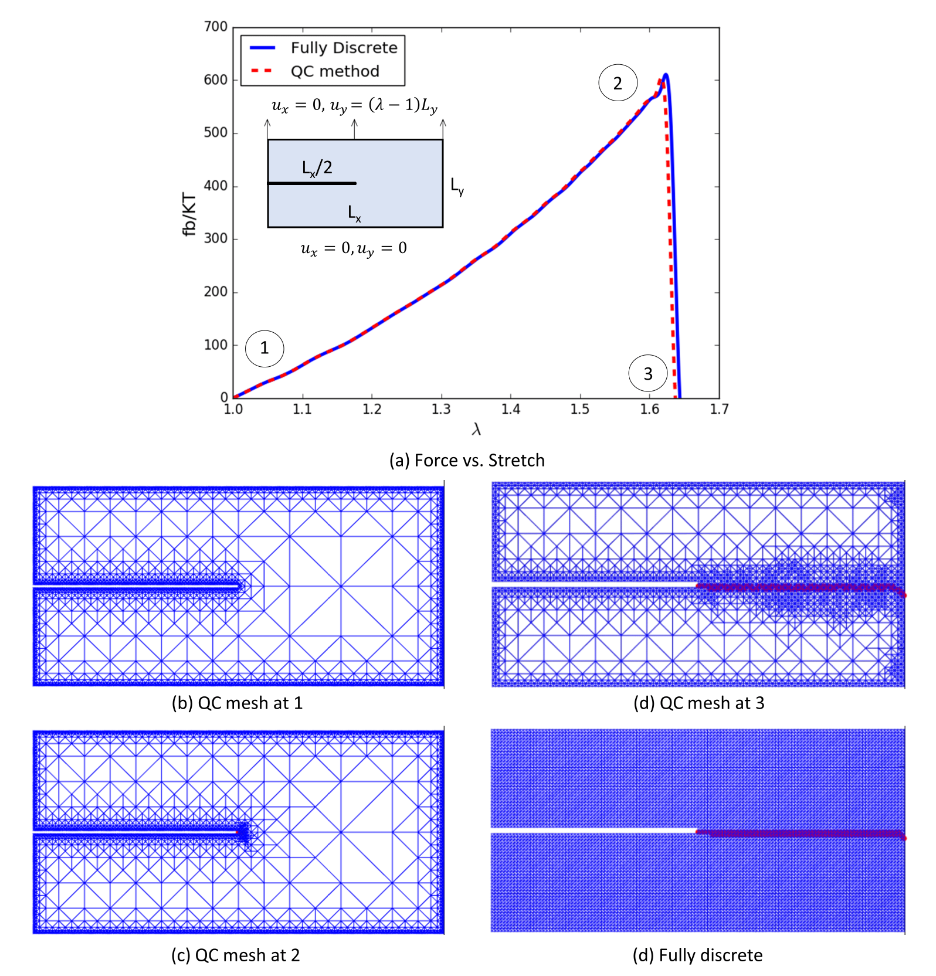}
\caption{Verification of the QC method for a fracture test: (a) the normalized force vs. stretch from QC and discrete results, the insert shows the geometry of the sample, (b,c,d) mesh evolution and crack path from the QC results, and (d) crack path from the discrete model results.}
\label{fig:Ver3}
\end{figure}

Here, we model crack propagation in a notched network using both fully discrete and QC models. The network dimensions are (500x250) chain units with around 250,000 degrees of freedom and the notch is 250 units long. The sample dimensions and boundary conditions are shown in the insert in Fig.\ref{fig:Ver3}-a. All links follow the worm like chain model, with force-stretch relation given by Eq.\ref{eq:pcf}. The chain length is drawn from a uniform random distribution with $\mu=0.05$. A link is set to break if the value of the stretch $\lambda_p$ exceeds 0.8. Automatic mesh adaptivty is turned on to allow the high interest area to evolve with the crack propagation.

The results in Fig.\ref{fig:Ver3} show that the crack path and the force vs. stretch curves are almost identical for both QC and fully discrete models. Specifically, the error in the peak force is around 1.0\%, the error in the peak stretch is 1.4\%, and the error in the total area under the curve is under 0.5\%. The QC mesh evolution is also shown in the same figure. The first mesh Fig.\ref{fig:Ver3}-b has a rep. nodes ratio of only 2.1\%, whereas the final mesh Fig.\ref{fig:Ver3}-d has 5\% rep. nodes ratio. This significant reduction in the number of degrees of freedom lead to extreme computational saving. 

\subsection{\textcolor{black}{Crack initiation:}}

\begin{figure}[H]
\centering\includegraphics[width=0.9\linewidth]{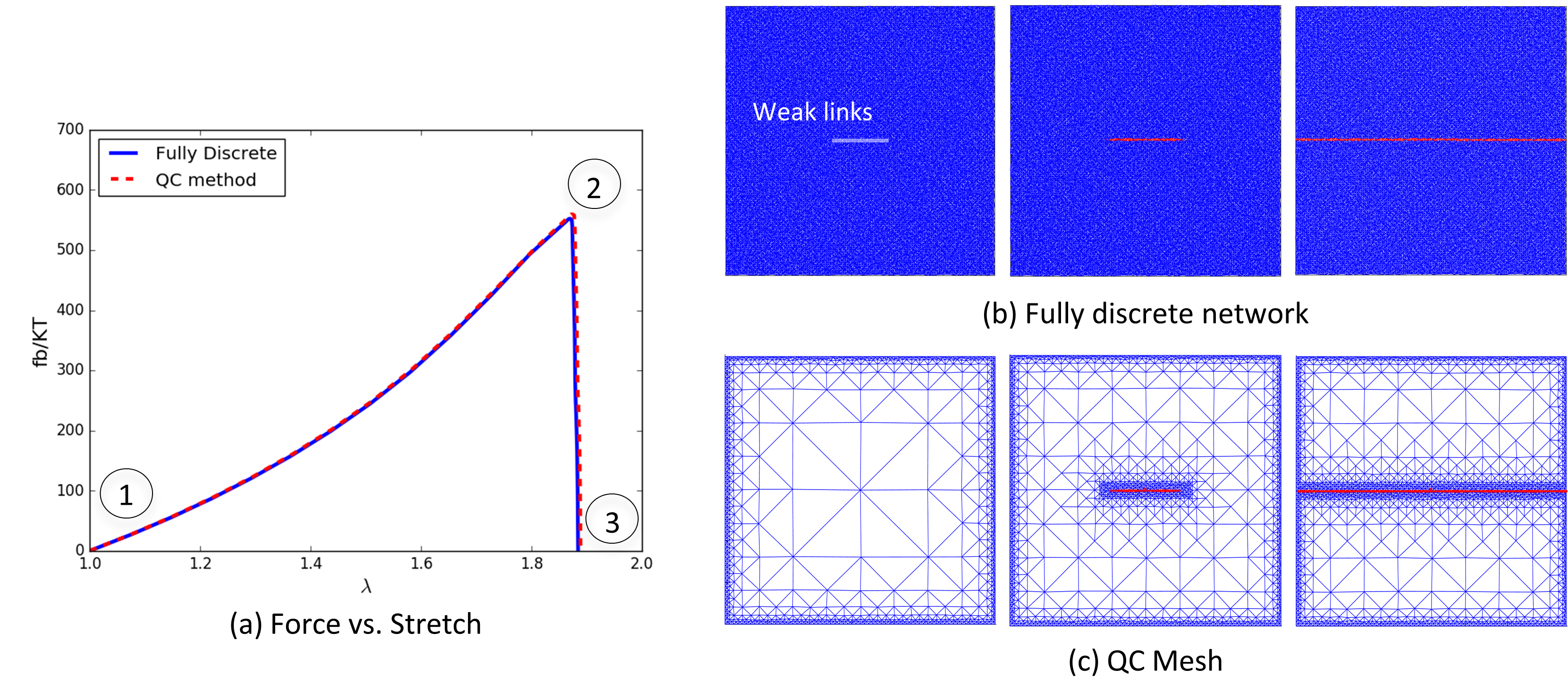}
\caption{\textcolor{black}{Crack initiation and propagation due to a zone with weak links: (a) The normalized force vs. stretch from QC and discrete results, (b) crack path from fully discrete simulation results at points 1, 2, and 3 respectively, and (c) mesh evolution and crack path from the QC results at points 1, 2, and 3 respectively.}}
\label{fig:CI}
\end{figure}

\textcolor{black}{In this section, we verify the capability of the proposed QC methodology to capture crack initiation in an intact sample and subsequent crack propagation. Formulating a criterion for crack nucleation in continuum models may be challenging since initiation of fracture in an intact sample may be based on strength rather than energetic consideration. As the QC blends both discrete and continuum representations, it does not suffer from such limitation. Fig.\ref{fig:CI} shows the comparison between the results of a fully discrete simulation and the corresponding QC mesh for a pristine sample with dimensions (250x250) where the links in the zone shown in Fig.\ref{fig:CI}-a have been weakened by lowering the failure stretch to 50\% of that of the network links. The QC method captures the crack initiation thanks to the link stretch based mesh refinement criterion. The force-stretch relation for both cases are nearly identical with errors in peak force and stretch below  0.5\%.}

\subsection{\textcolor{black}{Computational time analysis:}}
\textcolor{black}{In this section, we compare the run time for the proposed QC approach vs. that of fully discrete samples for the same geometry in Fig.\ref{fig:Ver3} and different sample sizes. For solving the system, we use conjugate gradient solver with incomplete Cholesky preconditioning \cite{Heath1997SCIENTIFICSurvey} which is suitable for symmetric positive definite matrices. The computational times of an average step and the total time of a simulation of 500 steps are shown in Table \ref{tbl:compu cost}. The time portions consumed in solving the system of equations at each step, including all iterations, are shown separately. Although QC requires more time at the initialization to construct the course mesh and assign links to QC elements, the total simulation time significantly reduces compared to the fully discrete simulations. The reduction is due to solving smaller system each time step, the time consumed in solving the system is reduced by nearly two orders of magnitude in the cases shown in Table \ref{tbl:compu cost}. In addition, the time consumed in the homogenization is less than the time required to calculate the local stiffness matrix for each link and assemble the global stiffness matrix in the fully discrete simulation. Thus, the total simulation time is reduced up to 15 times for the cases shown. Table \ref{tbl:compu cost} also shows that as the sample size increases, the ratio of the representative nodes decreases and the speed up ratio between QC and fully discrete simulation increases which suggests that computational advantage of the method is even larger for larger scale problems. It is also worth noting that homogenization step is done independently for each QC element, thus it may be easily parallelized with almost ideal expected speed-up \cite{Mikes2017QuasicontinuumLatticesb}.}

\begin{table}[H]
  \caption{\textcolor{black}{Computational times consumed by fully discrete vs QC method for different sample sizes}}
  \label{tbl:compu cost}
  \centering\includegraphics[width=1.0\linewidth]{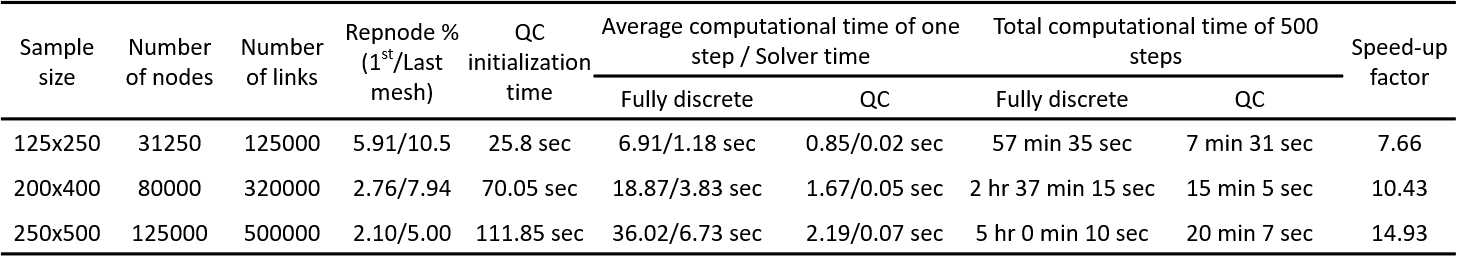}
\end{table}

\section{Results}\label{sec:network}
In this section, we apply the QC method to investigate the effects of local topology and disorder on the network mechanical properties and its fracture characteristics. We show the advantage of the QC method in capturing those effects without the need to explicitly model every node of the network. As explained earlier, the discrete representation is limited to the zone around the crack whereas the regions away from the crack are homogenized consistently. This introduces a computationally efficient accurate  representation of the fracture process. 

\subsection{Damage evolution in samples with different disorder parameter:}

\begin{figure}[H]
\centering\includegraphics[width=1.0\linewidth]{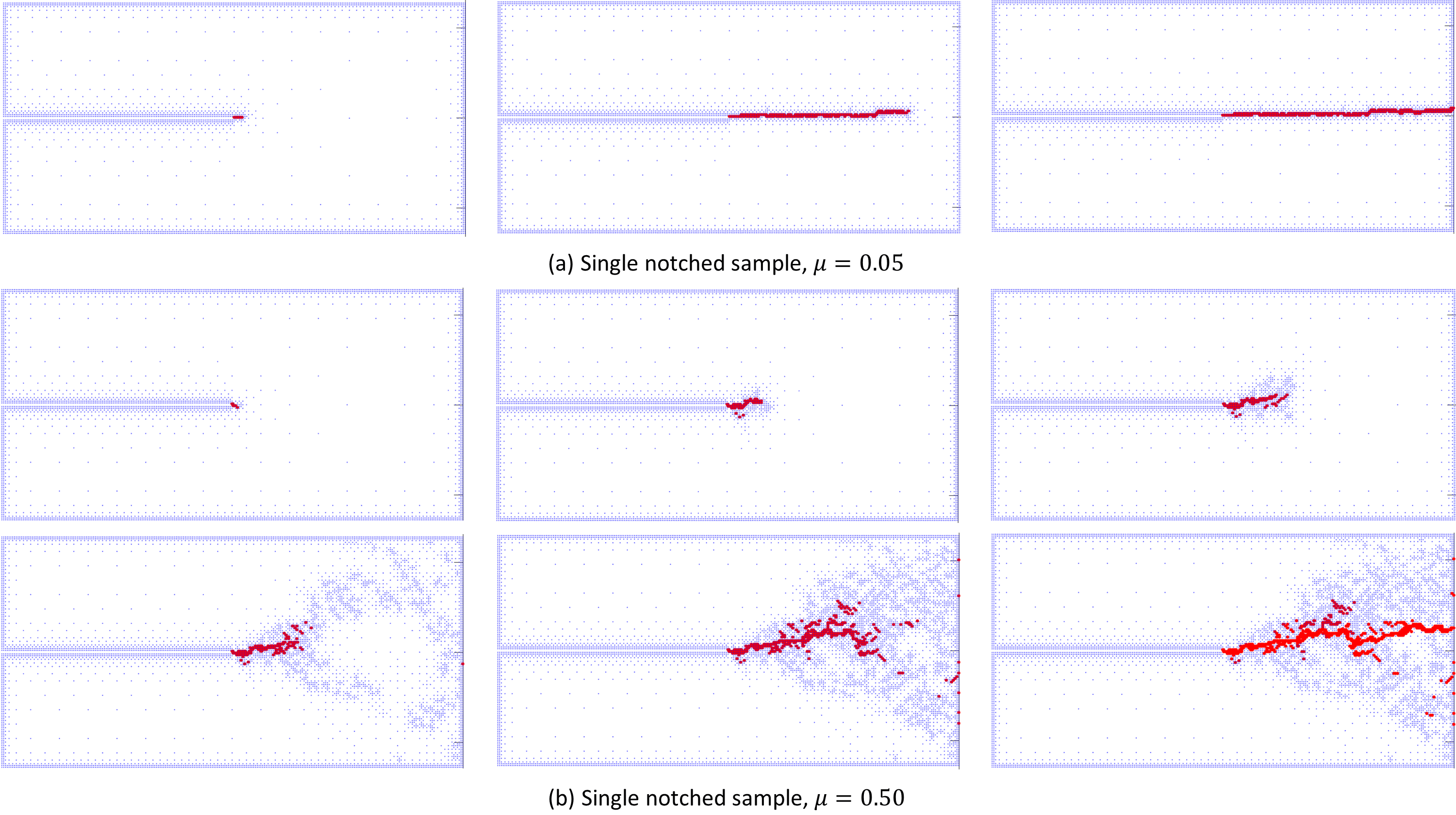}
\caption{Damage evolution in a single-notched sample, the sample geometry is shown in the insert of Fig.\ref{fig:Ver3}-a: (a) network with low disorder, and (b) network with high disorder. The damaged lattices are marked by red color, whereas the repnodes are shown as blue dots.}
\label{fig:Fail}
\end{figure}

Figure.\ref{fig:Fail} shows the damage evolution and the crack path for the single notched sample setup shown in Fig.\ref{fig:Ver3}-a with two different disorder parameter ($\mu$) values and $Z\approx 8$, where $Z$ is the coordination number defined as the number of connections a node may have to other nodes in the network, averaged over all the nodes. 

For $\mu=0.05$, which represents an almost uniform network, the crack propagates coherently perpendicular to the applied loading direction. The crack path is nearly horizontal, and the damage is localized along the crack path. For highly disordered networks with $\mu=0.50$, although the dominant crack propagation direction is, on average, perpendicular to the applied loading direction, the crack path has more kinks as it propagates. Furthermore, the crack thickness, which may be taken as a basis for a material length scale in nonlocal theories such as phase field methods, is not set apriori. It emerges as part of the solution and shows non-uniformity as it tends to increase at some locations and decreases in others. Damage is not localized around the crack path, as the figure shows multiple local damage zones away from the crack tip. Some of these local damage spots coalesce with the main crack while some remain isolated. Fig.\ref{fig:Fail} also shows the increase in the number of repnodes, represented by the blue dots, with crack propagation due to mesh adaptivity. The rate of increase in the number of repnodes for more disordered network is higher than that in the more uniform ones. This attests to the flexibility of the QC method in adapting to the complexity of the underlying network structure.

\subsection{Force distribution in the vicinity of the crack tip:}
To gain further insight into the processes controlling crack initiation and growth, we analyze the force distribution in the discrete network in the vicinity of the crack tip for different values of the coordination number $Z$ and disorder parameter $\mu$. As defined earlier, the coordination number is the average number of nodes directly connected to a given node and measures how densely connected the network is. 

\begin{figure}[H]
\centering\includegraphics[width=0.8\linewidth]{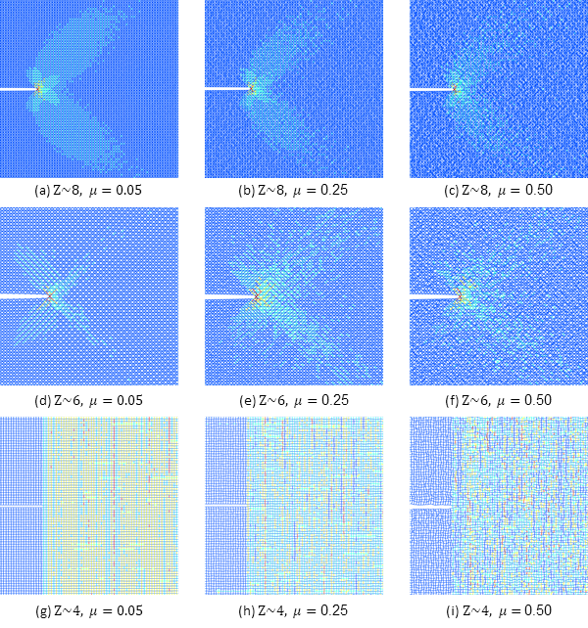}
\caption{The force distribution in the network links around the crack tip: (a,b,c) networks with $Z \approx 8$ and $\mu = 0.05 \text{, } 0.25 \text{, and } 0.5$ respectively, (d,e,f) networks with $Z \approx 6$ and $\mu = 0.05 \text{, } 0.25 \text{, and } 0.5$ respectively, and (g,h,i) networks with $Z \approx 4$ and $\mu = 0.05 \text{, } 0.25 \text{, and } 0.5$ respectively.}
\label{fig:SD}
\end{figure}

The results in Fig.\ref{fig:SD} show how changing the local connectivity changes the force field around the crack altering conditions for the crack initiation, propagation, and growth direction. For example, for $Z \approx 8$ and $\mu=5\%$, the stress field has the classical peanut-like distribution as expected for Mode I fracture in planar elasticity \cite{anderson2017fracture}. On the other hand, for $Z \approx 4$ and $\mu=5\%$, the anisotropic network channel the forces in the links that are closely aligned with the loading direction. For $Z \approx 6$, the force distribution shows an interesting pattern with some increase in the forces along diagonal directions running backward relative to the crack tip. For each coordination number, increasing the disorder in the chain length significantly changes the force field. In particular, the force field becomes more diffuse and loses its symmetry. Furthermore, increased disorder may lead to elevated force magnitudes in regions away from the crack tip leading to a more complex fracture pattern and potentially thicker cracks as was observed in Fig.\ref{fig:Fail}. Such complex force field distribution in highly anisotropic and disordered networks suggest the possibility of potential crack growth in unusual directions as has been observed recently as sideway cracks in silicone elastomers \cite{Lee2019SidewaysElastomer}. The QC methodology may enable systematic discovery of these new patterns.

\subsection{Effect of disorder in the chain length:}
\begin{figure}[ht]
\centering\includegraphics[width=0.8\linewidth]{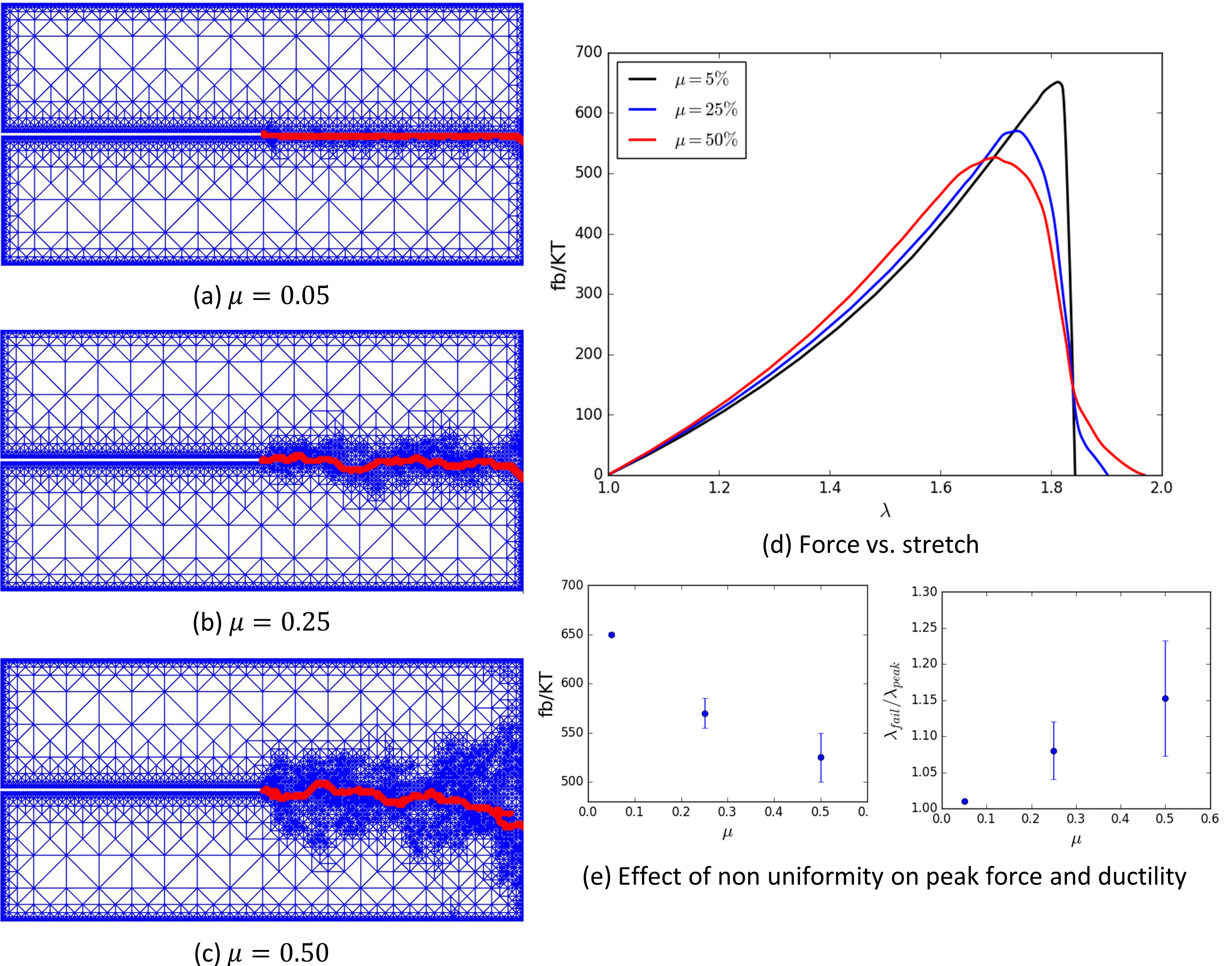}
\caption{\textcolor{black}{The effect of network disorder on the mechanical response of single-notched samples. The sample geometry is shown in the insert of Fig.\ref{fig:Ver3}-a: (a,b,c) crack path for single notched sample and 3 different disorder parameter $\mu = 0.05 \text{, } 0.25 \text{, and } 0.5$ respectively, (d) the force vs. stretch relation for the three disorder parameter values, and (e) the effect of $\mu$ on the peak force and ductility of the networks, the uncertainty in these values due to the disorder is indicated by the error bars.}}
\label{fig:SN}
\end{figure}

We simulate the disorder in the network by drawing the chain length from a stochastic distribution. Fig.\ref{fig:SN} shows the response  of a network with a single notch, the same setup shown in Fig.\ref{fig:Ver3}-a, when the available chain length is drawn from a random uniform distribution ranging between $[(1-\mu)L_c,(1+\mu)L_c]$ where $L_c$ is a predefined contour length. Fig.\ref{fig:SN}-a-c show the effect of the chain length variability on the crack path. As the variability increases, the crack path becomes more meandering while propagating through the network. The QC mesh also shows higher refinement (high repnodes ratio) as the disorder increases. This is because as the range of variability in the chain length increases,  some links with shorter contour length become critically stressed at locations not at the crack tip requiring further refinement in these regions. 

\begin{figure}[ht]
\centering\includegraphics[width=0.8\linewidth]{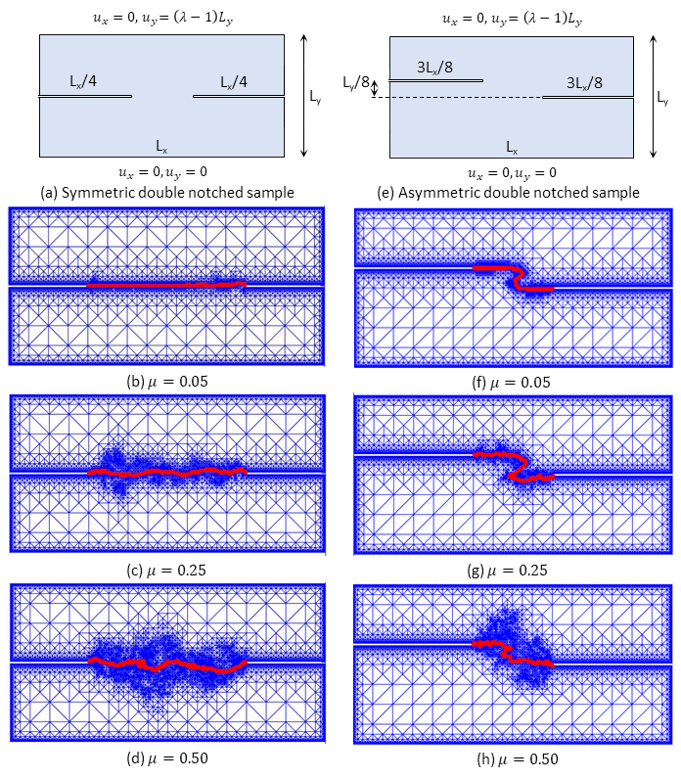}
\caption{The effect of network disorder on the crack path of double-notched samples: (a) the geometry of symmetric double notched sample, (b,c,d) crack path for symmetric double notched sample and 3 different disorder parameter $\mu = 0.05 \text{, } 0.25 \text{, and } 0.5$ respectively, (e) the geometry of asymmetric double notched sample, and (f,g,h) crack path for asymmetric double notched sample and 3 different disorder parameter $\mu = 0.05 \text{, } 0.25 \text{, and } 0.5$ respectively.}
\label{fig:DN}
\end{figure}

To further investigate the effect of disorder, the average force vs. stretch curves are plotted for different disorder parameters $\mu$ in Fig.\ref{fig:SN}-d. \textcolor{black}{Each curve is the average of number of randomly generated networks with the same disorder parameter. The number of networks was chosen such that the change in both the mean and the standard deviation of the peak force distribution due to adding more networks falls below 1\%. We performed this analysis for the highest disorder parameter which required 60 networks, then the same number of networks was used for all cases. More details about this procedure may be found in Appendix A.} The results show an increase in the stiffness and a decrease in the peak force as the disorder increases. The increase in stiffness is intrinsic in the nature of force stretch curve of the single worm like chain where the force increases rapidly as the stretch ratio approaches the stretch failure threshold. For more disordered networks, the increase in the force in the shorter links dominate over the reduction in force in the longer links. For the same reason, networks with higher variability in chain length tends to initiate failure at lower peak force, as they have a larger number of shorter links which fail faster. The results also show that the ductility (i.e. the ratio between stretch at total failure and stretch at peak force) increases with increasing $\mu$. In less disordered networks, the links are more uniformly stressed and hence they fail rapidly once failure initiates. As the variability in the chain length increases, the network has more longer chains which helps the material system to survive longer under damage, this agrees with the experimental results reported in Bonn et al. \cite{Bonn2014DelayedSolid}. Fig.\ref{fig:SN}-e shows that the average peak force decreases with increasing $\mu$, whereas the average ductility increases. Furthermore, the uncertainty in both peak force and ductility, represented by error bars, increase with increasing the material disorder. 

Figure.\ref{fig:DN} shows the effect of increasing the network disorder on the crack path of a network with symmetric and asymmetric double edge notches. As the disorder increases, crack path tends to be more nonuniform and thicker. In addition, a more extended spatial region is refined as the variability in the chain length increases, to capture the local failure of shorter chains in locations away from the crack tip. Furthermore, these results show how the two cracks emerging from the double notches interact with each other at different $\mu$ values. \textcolor{black}{The crack path shape for the case of $\mu=0.05$ was verified by running fully discrete simulations for the same geometry.} The fracture pattern agrees qualitatively with previously reported experimental and numerical observations \cite{Ghelichi2015ModelingMedia,Fender2010UniversalCracks} .It also agrees qualitatively with the numerical results for elastomeric materials fractured by cross-link failure using the phase field method \cite{Mao2018FractureFailure}. However, our results suggests that local topology may have a profound effect on this crack path that may be challenging to capture using fully homogenized approaches. In particular, the curving of the cracks towards one another may completely disappear as the disorder increases.

\subsection{Effect of coordination number Z:} \label{sec:Z}
\begin{figure}[H]
\centering\includegraphics[width=0.9\linewidth]{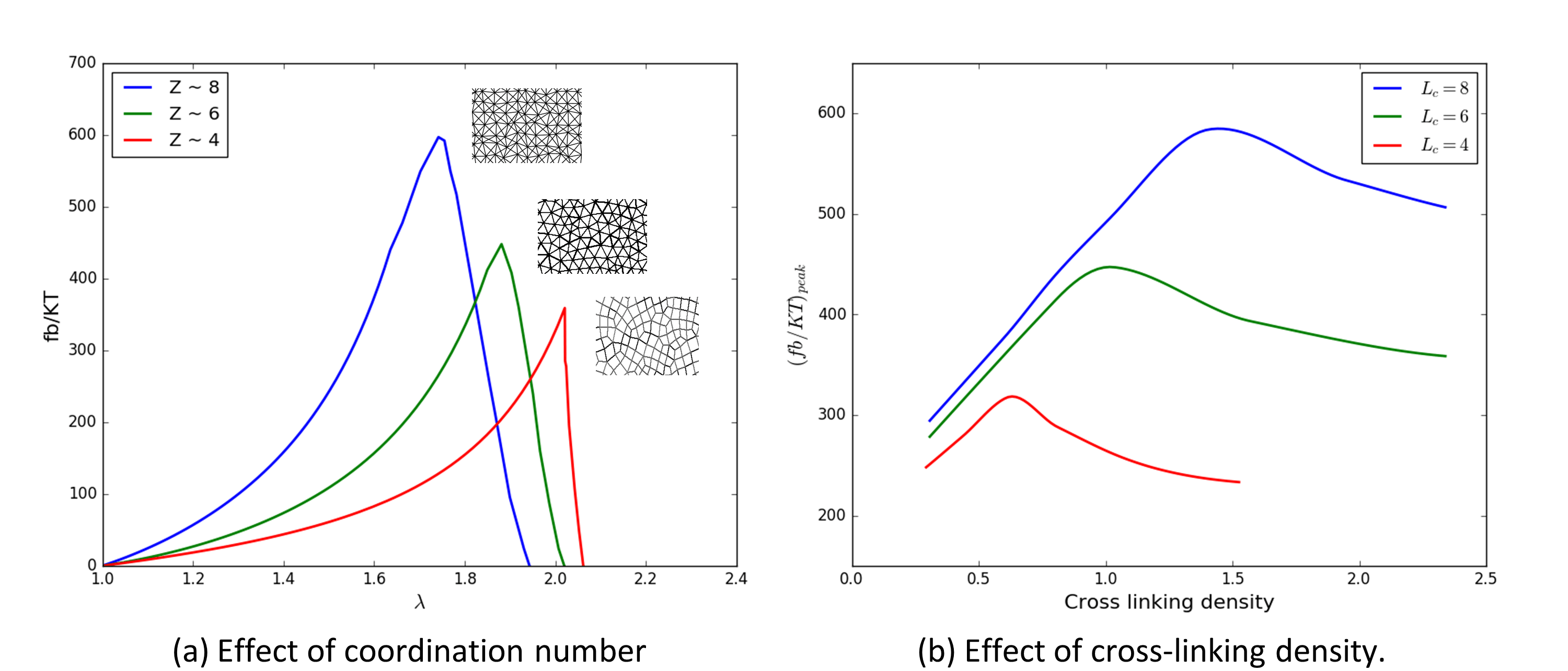}
\caption{\textcolor{black}{Effect of network topology on its mechanical response: (a) the effect of the coordination number $Z$ on the force vs. stretch behavior, and (b) the effect of cross-linking density, measured relative to a reference mesh, on the network peak force for three values of polymer chain available contour length $L_c$.}}
\label{fig:Z}
\end{figure}

In this section, we investigate the effect of network connectivity through changing he coordination number $Z$. Fig.\ref{fig:Z} shows the force vs. stretch for three different coordination numbers,  $Z\approx 4$, $Z\approx 6$ and $Z\approx 8$. In constructing these networks, we keep the total weight of the network ,i.e. the sum of all polymer chain average contour lengths, as a constant. We also draw the contour length of the different chains from the same uniform random distribution with $\mu=0.25$ for the three cases. The results show that networks with higher coordination number has higher toughness and peak force compared to networks with lower coordination  number. As the coordination number increases, more links contribute to load sharing and hence the network tends to become stiffer and sustain higher load. \textcolor{black}{Similar conclusion was recently reported by Alame and Brassart \cite{Alame2019RelativeNetworks}}. Furthermore, the case with higher coordination number shows  a larger ductility ratio, defined as the ratio between stretch at failure and the stretch at peak force, but a slightly smaller maximum stretch. As the coordination number increases, there are, on average, more links connected per node. This enables a more efficient force redistribution among neighboring chains when one of the chains break off, leading to slower force drop and delayed damage accumulation. Similar results were obtained for nano-composite gels with nanoparticle cross-linkers that increase the network coordination interactions \cite{Wang2016ACrosslinkers}. As the coordination number decreases, the material becomes more compliant, and is capable of achieving slightly higher stretch values although with lower overall toughness and strength. Specifically, the area under the curve for $Z\approx 6$ is 1.5 times that for $Z\approx 4$ and the area under the curve for $Z\approx 8$ is twice that for $Z\approx 4$.

\subsection{Effect of cross-linking density:}
To investigate the effect of increasing cross-linking density, defined as the number of network nodes per unit area, we simulate multiple samples with increased number of nodes and links in the the network while keeping the polymer concentration constant. This is achieved by adjusting the average chain contour length such that the sum of lengths of all chains remains constant. The results in Fig.\ref{fig:Z}-b shows that the peak force increases with increasing the cross-linking density up to a maximum value, then it starts decreasing. This non monotonic response is due to the reduction of the average chain length with increasing the cross-linker concentration. When the average chain length decreases, the network becomes stiffer and attains higher forces for a given value of stretch. However, as the cross-linker concentration further increases beyond a certain value, which depends on the chain length, the network becomes critically stressed as the initial stretch of the short chains becomes closer to their contour length. These short chains accumulate damage at a faster rate leading to a premature failure at lower global force levels. With increasing the available chain length, the peak force for a given concentration increases, and the critical cross-linker concentration corresponding to this peak force also increases. 

\section{Discussion:}\label{sec:discussion}

\textcolor{black}{Fracture is a highly nonlinear problem in which macroscopic response depends on the details of microscopic physics. Here we have introduced an adaptive model based on the Quasi-Continuum approach for analysis and design of networked materials that starts from fundamental link scale mechanics and consistently computes macroscopic response. While this work builds on recent developments of extending the quasi-continuum approaches to lattice materials, it brings new innovations in both the application domain as well as the method formulation. For example, on the application level, a major contribution of this study is to bring the QC method into the discussion of modeling fracture in polymeric and soft materials. This is a topic of intense contemporary interest which lends itself naturally to the QC framework but has been addressed in the solid mechanics community, for the most part, using continuum models or phase field approaches that generally overlook the underlying topological structure or have limited predictive power. Furthermore, a unique advantage of the QC methodology is that it allows capturing crack initiation and nucleation, a strength-based process, that other approaches built on modified Griffith criterion, may fail to capture in a straightforward way. On the method formulation, due to the highly nonlinear nature of the polymer networks, we have proposed a homogenization rule for axially loaded lattices that accounts for both material and geometric nonlinearities and implemented it in a nonlinear finite element framework.  Moreover, we have developed an adaptivity scheme that allows for transition between homogenized material and network lattices where the refinement criteria depend not only on the deformation gradient, as has been done in earlier studies but also on link stretch. This is necessary, particularly for highly disordered networks, and also to allow for triggering refinement and crack nucleation at the locations of weaker links in absence of a notch or pre-crack }

While methods for modeling the nonlinear elastic response of polymeric materials are well established \cite{Boyce2000ConstitutiveReview,M.Arruda1993AMaterials} , modeling fracture and damage in soft polymeric materials continue to elicit challenges in science and engineering  \cite{Qi2018FractureHysteresis,Zhao2012ANetworks,Nguyen2005AMechanics,Guo2018FractureStudy,Yu2018MechanicsBonds,Talamini2018ProgressivePolymers,Mao2018AGels,Poulain2017DamageMacro-cracks,Lavoie2019AElastomers}. However, despite enabling useful insights, most of the existing techniques suffer from a number of limitations. For example, most continuum fracture models are insensitive to local topological or micro-structural details at the network level. They also require input for critical parameters such as fracture energy which may not be easily measured in many soft materials or may exhibit size effects \cite{Zhang2019Y-ShapedTearing}. For phase field methods, and other non-local theories, a length scale is also needed for regularization of the fracture \cite{Pijaudier-Cabot1988NonlocalTheory}. While in the case of phase field approaches, this length scale is introduced for numerical purposes and  the framework should converge to Griffith formulation upon taking this length scale to the zero limit, it is not clear how this length scale is to be chosen in cases when the material has an intrinsic one arising from its micro-structure. This compromises the predictive power of these modeling approaches. While fully discrete simulations provide a powerful alternative, as they capture the full details of the material microstructure, the computational cost of these approaches become prohibitive as the sample size increases. 

The main advantage of the QC framework is that it enables efficient modeling of large scale problems without neglecting the effect of local topology and without assuming any critical macroscopic parameters such as fracture energy, or material length scale. Only a local failure criterion at the chain level is needed and this may be rigorously computed directly from explicit model of a single chain \cite{Lieou2013SacrificialBone,Kothari2018MechanicalAccumulation,Mao2018FractureFailure}. The QC method requires much less run time, and memory than a full discrete model while providing higher accuracy and insights into the fracture process than the continuum approaches. Furthermore, the QC has the flexibility of adding small scale physics and topology. This makes it suitable for taking advantage of recent advances in experimental techniques. 

While most current experimental techniques do not allow direct visualization of the deformation and damage on the network micro scale, there have been a few exciting advances in recent years towards that goal. For example, the work of Ducrot et al. \cite{Ducrot2014TougheningBreak}  where a special type of mechanophore molecules is embedded in the polymer network and enables the direct imaging of the spatio-temporal evolution of damage density due to bond breakage. Also the work of Poulain et al. \cite{Poulain2017DamageMacro-cracks} where high speed photography is used to track the onset and growth of cavitation and subsequent coalescence and crack growth. These recent advances in imaging and experimental measurements among others are probing deformation and failure mechanisms in soft materials with increasingly high resolution challenging some long-held ideas about these phenomena \cite{Lefevre2015CavitationPhenomenon,Millereau2019DetectionMechanophores}. The proposed QC approach is uniquely suited for taking advantage of these advances and making use of the increasingly available data at the microscale. Furthermore, the QC approach provides a rigorous tool for linking the micro and macroscales with minimal set of assumptions which may potentially help resolve several challenges in failure of hierarchical and heterogeneous polymeric materials. 

It is worth to note that there exists a large literature on discrete models of fracture that abstract a continuum domain using a lattice representation, such as Lattice Discrete Particle Modeling \cite{Cusatis2011LatticeTheory}, Graph-based Finite Element Analysis GraFEA \cite{Khodabakhshi2016GraFEA:Materials}, and Virtual bond model \cite{Zhang2013DiscretizedElasticity}. A major difference between the QC approach and these methods is that the discrete component of the QC model is physical and represents an actual networked material with a well defined failure criterion at the discrete elements scale. Thus, the QC method does not suffer from some limitations that exist in these other lattice models such as mesh sensitivity, possible ambiguity in defining the fracture criteria upon abstraction of the continuum scale to the lattice scale, or the necessity of introducing non-local effects \cite{Pijaudier-Cabot1988NonlocalTheory} or phenomenological length scales. In particular, the crack path computed from the QC model is not mesh dependent, addressing an important challenge in modeling fracture of inhomogeneous materials \cite{Bazant2010CanFailure}. 

Our results for fracture of disordered networks suggest that as the disorder increases, the expected value of the peak force decreases but the ductility ratio and energy dissipation increase. It is also observed that the crack path becomes nonuniform with increased disorder, as the fracture mechanism includes numerous links failing in the network away from the crack tip zone and later coalescing into a thick crack path (Fig.\ref{fig:Fail}-b). Meandering crack paths suggest a more ductile response, which is observed in Fig.\ref{fig:SN} -d,e and increases fracture toughness compared to straight cracks. Such rough cracks are observed extensively in experiments \cite{Ravi-Chandar1984AnBranching,Carlson2018Watch3-D,2015APS..DFD.A1004S} and are usually attributed to dynamic instabilities. Here we show that they may also emerge naturally due to the increased level of disorder in the network structure. Transition in the failure mode from brittle to ductile with increased disorder has been reported previously using fully discrete models on smaller samples \cite{Driscoll10813}. The QC method provides a flexible tool for reassessing these phenomena at larger scales and with less sensitivity to sample boundaries. 

The current formulation of the QC implementation has a few limitations. For example, we have assumed an affine projection approximation to couple micro and macro-scales in the homogenized region of the domain. While we have shown that the homogenization error is within acceptable limits for the cases tested, this affine approximation has some disadvantages. For instance, the accuracy was shown to deteriorate as the disorder increases unless mesh adaptivity is turned on. In addition, we only considered networks with axial interactions. Some polymer chain models have a contribution to their potential energy originating from the rotational degrees of freedom which introduces a non-affine component in the deformation field \cite{Broedersz2014ModelingNetworks}. While any force-stretch law may be readily used in the current formulation in place of the worm-like chain model, future work will explore extension of the homogenization algorithm to account for non-affine deformation.

\textcolor{black}{In this initial study we opted to evaluate the stress and material tangent tensors from the contributions of all links underlying each continuum elements. It is possible, however, to determine a representative volume element (RVE) and estimate the stress and tangent tensors from the contribution of links within this volume. The general form of the tensors would be the same, but only links within this RVE will be accounted for. While this homogenization approach leads to increased computational savings, it also has some drawbacks. For example, it is not clear apriori how large the RVE should be in the case of disordered networks. Also the approximation based on an RVE will introduce additional errors. The pros and cons of this alternative methodology will be investigated further in future work.} 

\textcolor{black}{Automatic mesh adaptivity leads to significant computational savings by reducing the initial fully resolved zone. However, as the crack propagates, the ratio of repnodes increases, especially in cases of highly non-uniform networks, leading to reducing the computational savings. It is possible to implement coarse-graining behind the crack process zone to reduce the number or repnodes when the full resolution of a specific zone is not further needed. QC with coarse graining was accomplished by Rokos et al \cite{Rokos2017EXtendedPropagation} were significant computational savings were reported. }\textcolor{black}{In addition, time adaptivity allows taking larger time steps, when possible, without compromising accuracy. In the current study, we use the same time step for the whole domain. The choice of time step is based on the smallest stable value possible in the global assembly. However, it is possible to adopt an asynchronous time stepping approach where different time steps for the fully discrete and homogenized domains are used and couple them in a parallel framework \cite{Astorino2016AMethods}. This might lead to significant time savings as the fully discrete domain usually requires small time steps to resolve the fracture accurately and will be explored further in a future study.} 
 
Viscoelasticity is a critical feature in the mechanical response of polymeric materials such as rubber and gel. In our prior work \cite{Lieou2013SacrificialBone,Kothari2018MechanicalAccumulation}, we have modeled discrete polymer networks with viscoelastic response emerging from two major contributions: (1) breakage and formation of end bonds and cross-linkers, and (2) vsicous drag arising from the relative motion of the deforming polymer network and infused fluids. Extensive work has also been done previously by other groups for experimental characterization of the elastic and loss mdodulii \cite{Li2012ExperimentalGels,Nguyen2010ModelingPolymers}. Continuum models based on phenomenological as well as homogenization approaches also exist \cite{Kumar2016OnMaterials,Zhang2015MesoscaleComposites}. The QC method is capable of incorporating the viscoelastic response. The key point is to map the viscoelastic behavior from the microscale (network level) to the macroscale (QC level) using a proper homogenization rule. As a starting point, one may assume a damping force in each link that is proportional to the link stretch rate in addition to damping force at each network node proportional to its velocity. Then this damping matrix may be homogenized in a similar way as the stiffness component. The first term will result in a damping matrix that has the same structure as the stiffness matrix whereas the latter will result in a diagonal damping matrix. We will then use predictor-corrector schemes to solve the resulting linearized quasi-dynamic system at each time step. This approach would lead to the emergence of viscoelasticity on the continuum scale.

\textcolor{black}{In this initial work, we considered polymer networks in two dimensions. However, the method may be easily extended to model networks in the 3D domain. Unlike in atomistic systems, where the QC computational cost may exponentially increase from 2D to 3D due to the presence of long range interactions, the extension in lattice systems is straightforward since physical links introduce interaction between directly connected nodes only. Furthermore, extension to 3D will enable incorporation of  additional physical mechanisms such as entanglement, which may be modeled at the discrete level as an additional type of connections between the links, such as ring sliding connections \cite{doi:10.1002/app.40509} .This may then be homogenized using similar methodology to the one presented in this work and passed on to the continuum scale.}

To summarize, the QC framework enables significant insight into the role of microstructure on macroscopic fracture response in networked materials. The unique potential of the method lies in its predictive power which qualifies it to play the role of a highly controlled virtual experimental setup. This will not only lead to identification of fundamental processing controlling fracture and deformation in a variety of networks, including polymers, but may also enable the discovery of new material designs with unusual fracture response.

\section{Conclusions:}

Our main conclusions are summarized as follows:
\begin{enumerate}
    \item QC method provides a promising platform for investigating fracture in disordered networked materials that combines advantages from both discrete and continuum approaches. 
    \item The QC method for fracture in polymer networks only requires a failure criterion on the chain level. It does not require information about critical energy release rate or a material length scale. The QC discover these quantities as part of the solution. This predictive power and scale bridging capabilities give the QC framework an advantage over other existing methdologies. 
    \item Adaptivity allows the QC mesh to evolve as the crack grows, increasing accuracy and computational savings. For the cases presented here, we have achieved \textcolor{black}{up to 15 times speed up} compared to the full discrete models. Ongoing work with parallel implementation will enable bridging a wider range of scales with further computational savings.
    \item Crack path is controlled by the network topology and stochastic properties of the chains. Thus, QC method illuminates new pathways for design through direct control of the microstructure.
    \item The details of the network local topology have a significant effect on the local force distribution around the crack tip. Disorder makes the crack tip force field more diffused and  also channel forces in directions dominated by network anisortropy. 
    \item Networks with higher disorder in the geometry or/and chain length have higher stiffness but lower average peak force. Disorder also enhances the softening behavior as disordered networks fails in a more gradual manner compared to uniform networks.
    \item For networks with the same polymer chain concentration, increasing the coordination number leads to stiffer networks with higher peak force, lower stretch at failure, and more gradual softening behavior. 
    \item The peak force has a non-monotonic dependance on  the cross-linker concentration where it increases first as the cross-linker concentration increases but then decreases if the concentration exceeds a threshold value. Networks with longer chains have higher threshold value than those with shorter ones. 
\end{enumerate}
\section*{Acknowledgment}
This work is supported by the Center for Geologic Storage of CO2, an Energy Frontier Research Center funded by the U.S. Department of Energy (DOE), Office of Science, Basic Energy Sciences (BES), under Award No. DE-SC0C12504, and the National Science Foundation CAREER Award (Grant No. 1753249).
\newpage
\section*{\textcolor{black}{Appendix A: Effect of disorder on the force stretch relation}}

\textcolor{black}{We perform a simple statistical analysis to determine the minimum number of samples enough to draw some representative conclusions about the effect of network disorder on the force-stretch relation. We start with the highest disorder parameter ($\mu$=0.50), and we determine the number of simulations based on the saturation of the mean and the standard deviation of the peak force distribution where the rate of change of both the mean the standard deviation after adding additional runs falls below 1\%. For the specific parameters we use, saturation is achieved after around 60 random networks. We used this number for all other disorder parameters for consistency although lower number is sufficient in these cases.}

\textcolor{black}{Fig.\ref{fig:stat} a-c show the normalized force vs stretch curves for the 60 runs along with the average normalized force-stretch curve for each disorder parameter. Fig.\ref{fig:stat} d,e show the evolution of the mean and standard deviation with the number of runs. The average plots are shown in the manuscript.}

\renewcommand\thefigure{A.\arabic{figure}}    
\setcounter{figure}{0}
\begin{figure}[H]
\centering\includegraphics[width=1.0\linewidth]{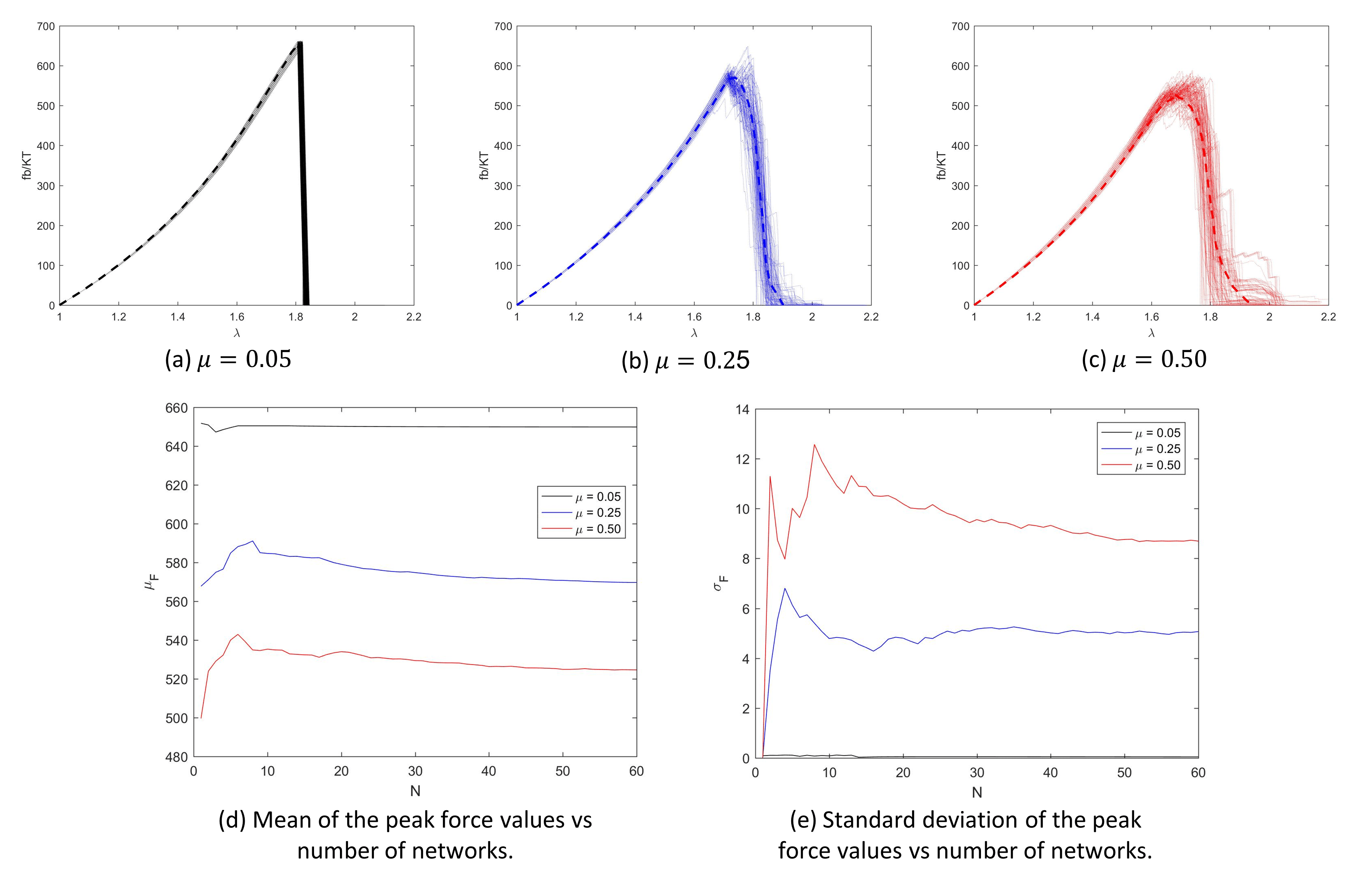}
\caption{\textcolor{black}{The results from the statistical analysis of disorder networks: (a-c) Normalized force vs, stretch curves for all simulated networks with disorder parameters of  $\mu = 0.05 \text{, } 0.25 \text{, and } 0.5$ respectively. The average curves are also plotted for each case, and (d,e) The evolution of the mean and the standard deviation of the peak force with the number of simulated networks for the three different studied disorder parameters.}}
\label{fig:stat}
\end{figure}

\pagebreak
\bibliographystyle{model1-num-names}
\bibliography{Mendely.bib}

\end{document}